\documentclass[a4paper,fleqn,usenatbib]{mnras}
\DeclareSymbolFont{matha}{OML}{txmi}{m}{it}
\DeclareMathSymbol{\varv}{\mathord}{matha}{118}
\newcommand\rs[1]{_\mathrm{#1}}

\usepackage{mathptmx}

\usepackage[T1]{fontenc}
\usepackage{ae,aecompl}

\usepackage{graphicx}	
\usepackage{amsmath}	
\usepackage{amssymb}	
\usepackage{bm}

\title[Radio evolution of young SNR G1.9+0.3]{Hydrodynamical and radio evolution of young supernova remnant G1.9+0.3 based on the model of diffusive shock acceleration}

\author[M. Z. Pavlovi\'c]{M. Z. Pavlovi\'c$^{1}$\thanks{Contact e-mail: \href{mailto:marko@math.rs}{marko@math.rs}}
\\
$^{1}$ Department of Astronomy, Faculty of Mathematics, University of Belgrade, Studentski trg 16, 11000 Belgrade, Serbia
\\
}

\date{Accepted 2017 February 23. Received 2017 February 9; in original form 2016 December 28}

\pubyear{2017}

\begin{document}
\label{firstpage}
\pagerange{\pageref{firstpage}--\pageref{lastpage}}
\maketitle

\begin{abstract}
The radio evolution of, so far the youngest known, Galactic supernova remnant (SNR) G1.9+0.3 is investigated by using three-dimensional (3D) hydrodynamic modelling and non-linear kinetic theory of cosmic ray (CR) acceleration in SNRs. We include consistent numerical treatment of magnetic field amplification (MFA) due to resonant streaming instability. Under the assumption that SNR G1.9+0.3 is the result of a Type Ia supernova explosion located near the Galactic Centre, using widely accepted values for explosion energy 10$^{51}$ erg and ejecta mass 1.4 $M_{\sun}$, the non-thermal continuum radio emission is calculated. The main purpose of this paper is to explain radio flux brightening measured over recent decades and also predict its future temporal evolution. We estimate that the SNR is now $\sim$ 120 yr old, expanding in an ambient density of 0.02 cm$^{-3}$, and explain its steep radio spectral index only by means of efficient non-linear diffusive shock acceleration (NLDSA). We also make comparison between simulations and observations of this young SNR, in order to test the models and assumptions suggested. Our model prediction of a radio flux density increase of $\sim$ 1.8 per cent yr$^{-1}$ during the past two decades agrees well with the measured values. We synthesize the synchrotron spectrum from radio to X-ray and it fits well the Very Large Array, Molonglo Observatory Synthesis Telescope, Effelsberg, \emph{Chandra} and NuSTAR measurements. We also propose a simplified evolutionary model of the SNR in gamma rays and suggest it may be a promising target for gamma-ray observations at TeV energies with the future generation of instruments like Cherenkov Telescope Array. SNR G1.9+0.3 is the only known Galactic SNR with the increasing flux density and we present here the prediction that the flux density will start to decrease approximately 500 yr from now. We conclude that this is a general property of SNRs in free expansion phase.
\end{abstract}




\section{Introduction}

\indent A potentially young shell-type Galactic supernova remnant (SNR)
G1.9+0.3 was identified for the first time by \citet{green84}, from Very Large Array (VLA) observations of a sample of small-diameter Galactic radio
sources. The interest in this SNR has increased after the work of \citet{reyn08} and \citet{green08} who deduced that G1.9+0.3 is of order 100 yr old and therefore, the youngest SNR in the Galaxy. Based on extremely high absorption, they placed G1.9+0.3 near the Galactic Centre (GC), at a distance of about 8.5 kpc, where the mean diameter would be about 4 pc and the required expansion speed about 14\,000 km/s \footnote{Derived from both expansion proper motions and Doppler shifts of lines from isolated regions of thermal emission.}. \citet{roy14} propose a lower limit on its distance from Sun as 10 kpc, based on the Giant Metrewave Radio Telescope (GMRT) measurements of absorption by known anomalous velocity features near the GC.

According to \citet{reyn08}, the synchrotron-dominated X-ray spectrum   clearly indicates that the effective particle acceleration takes place, at least for electrons, given the very high shock velocities and low ambient densities. Also, the implied characteristic roll-off electron energy of about 100 TeV is the highest ever reported for a shell SNR.

\citet{borkow13} reported spatially resolved spectroscopy of SN ejecta
and interpreted their results in the framework of an energetic and
asymmetric Type Ia explosion. They also concluded that the outermost ejecta layers in free expansion have velocities in excess of 18\,000 km/s. Several arguments suggest a Type Ia origin of G1.9+0.3 \citep[see also][]{reynolds08}: the high velocities more than 100 yr after the explosion, the absence of central pulsar-wind nebula and the bisymmetric morphology in X-ray and substantial thermal emission from Fe. A usual core-collapse event could not reproduce the observations, while an SN Ia model can easily reach the observed size and velocity for a mean ambient density of about 0.02 cm$^{-3}$ \citep{carlton11}.

\begin{figure}
	\includegraphics[width=1.04 \columnwidth]{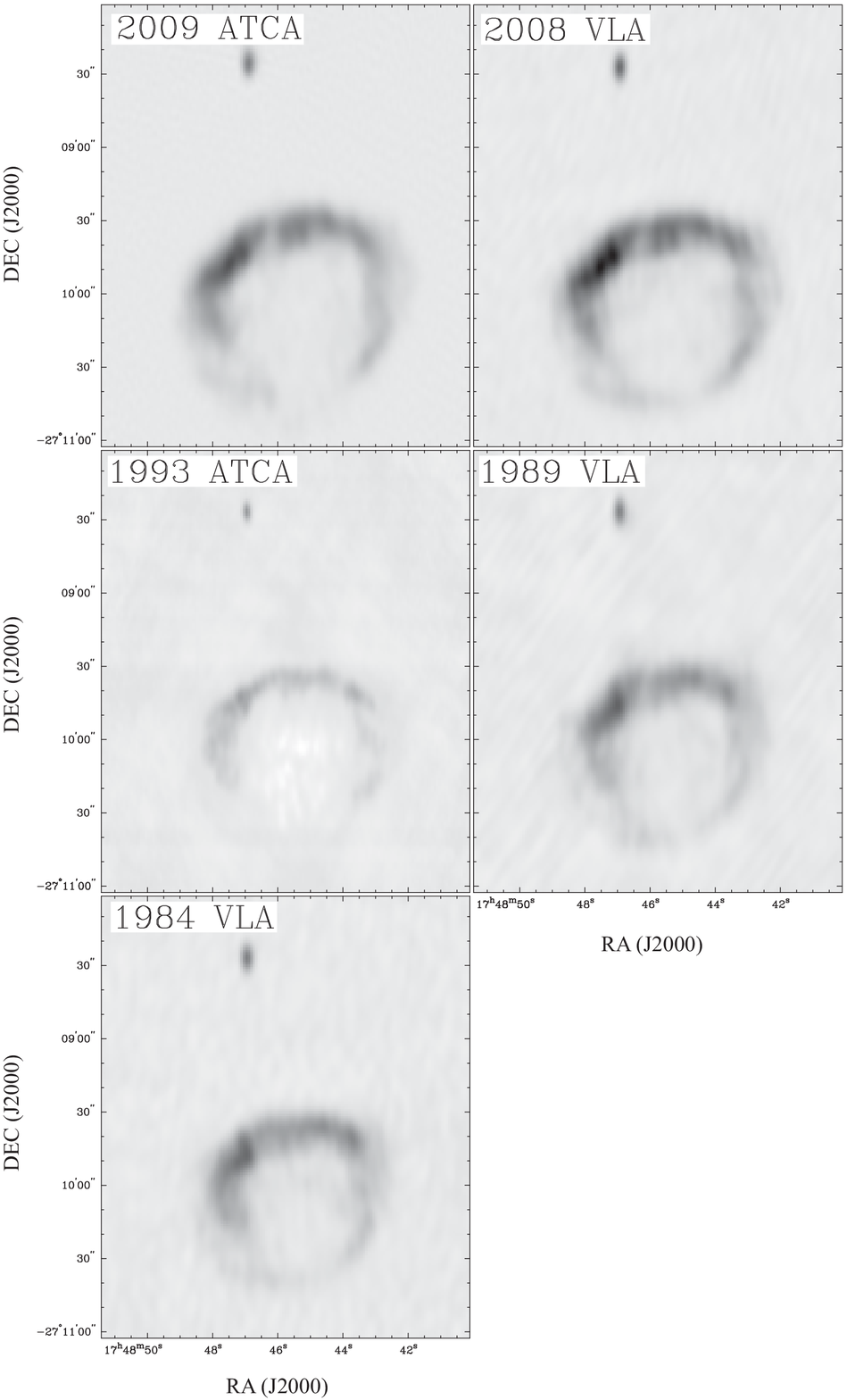}
    \caption{Matched resolution 6-cm images of Galactic SNR (centred at RA(J2000)=$17^\mathrm{h}48^\mathrm{m}45^\mathrm{s}.4$, Dec(J2000)=$-27\degr10\arcmin06\arcsec$) at multiple epochs. Left to right, top to bottom, 2009 ATCA, 2008 VLA, 1993 ATCA, 1989 VLA and 1984 VLA \citep{horta14}.}
    \label{fig:radio}
\end{figure}

\citet{green08} compared their VLA radio observations of the SNR G1.9+0.3 at 4.86 GHz and 1.43 GHz with earlier observations at 1.49 GHz which have a comparable resolution. They found evidence that this SNR has been brightening over the past few decades in the radio emission at a rate of $\approx$ 2 per cent yr$^{-1}$ for the available flux densities and proposed explanation that the efficiency of particle acceleration and/or magnetic field amplification (MFA) has been increasing. \citet{horta14} analysed all available radio-continuum observations of SNR 1.9+0.3 at 6 cm from the VLA and also the Australia Telescope Compact Array (ATCA) (see Fig. \ref{fig:radio}), obtaining results that are in broad agreement with the estimates of expansion made by \citet{reyn08} and \citet{green08}.

By using the time-dependent non-linear kinetic theory for cosmic ray (CR) acceleration in SNRs, coupled with 1D spherically symmetric gas dynamics, \citet{bif04} predicted that radio luminosity should increase during the free expansion phase, in general case. According to these authors, this is mainly due to the growing number of accelerated CRs. \citet{ksenofontov10} also applied a similar model of non-linear CR acceleration to study the non-thermal properties of SNR G1.9+0.3. They obtained a spatially integrated radio synchrotron flux that slowly increases with time, explaining it as a consequence of the rapidly increasing total number of accelerated electrons in the increasing SNR volume $\propto R_{\rm{s}}^{3}$, where $R_{\rm{s}}$ represents the current shock radius.

Recently, \citet{chak16} used their analytical model to demonstrate that a double degenerate (DD) progenitor can explain the decades-long flux rise and size increase of the SNR G1.9+0.3 and disfavour a single degenerate (SD) scenario for this SNR. Nevertheless, for the pre-explosion circumstellar density $\rho_{\rm cs} \propto r^{-s}$ they assume index $s=2$ for the SD and $s=0$ for the DD case, which may be questionable.

It should be examined whether the increasing radio brightness of G1.9+0.3 is a unique property amongst the SNRs in our Galaxy and does it require some special conditions. One of the aims will also be to go beyond the specific analysis of G1.9+0.3 and, in some future papers, expand the proposed analysis to any young SNR or further to a global sample of SNRs with different ages.

Why do we think the radio evolution modelling is important? Electrons emitting at a radio frequency have the acceleration time-scales of the order of a week, when the Bohm diffusion is assumed \citep{petruk16}. This is much less than the acceleration time of the highest-energy particles, for which it is of the order of an SNR age. Due to this, modelling of radio evolution and connecting it to the radio observations may reveal the
present-time behaviour of the injection efficiency and also its time
dependence \citep{petruk16}.

Also, the radio surface-brightness-to-diameter ($\Sigma-D$) relation for
SNRs, being a useful distance determination tool, can be significantly
improved if the radio evolution is better understood. This relation is known to depend on the properties of the SN explosion such as the explosion energy, mass of the ejected matter and also on the properties of ISM such as density, magnetic field strength, etc. One of the main shortcomings of this relation is the severe data scatter that is mainly due to the spread in mentioned parameters, in addition to measurement errors and selection effects \citep[see, e.g.][etc.]{arbo05, urosevic10, pavlovic14}.
Theoretical considerations contain many limitations because they often rely on simplified assumptions about the evolutionary stage of SNRs, particle spectra and its evolution, magnetic field evolution, etc. Numerical simulations should provide a better understanding of underlying physics and explanation of the observed statistical properties.

Magnetic field is one of the main ingredients in particle acceleration and non-thermal emission. Consistent, the time-dependent calculation of MFA is one of the main advantages in an approach based on numerical simulations.

\section{Model}
The dynamical evolution of an SNR was modelled by numerically solving the time-dependent hydrodynamical (HD) equations of mass, momentum and energy conservation, including a semi-analytical model of acceleration and back reaction of particles on shock dynamics. We threat the back-reaction of the energetic particles on the shock in sense of the pressure of particles that affect shock dynamics \citep{blasi04} and CR current that amplifies the magnetic field \citep{bell04, caprioli09}. Both effects happen upstream of the shock wave, in the so-called precursor. Accent of this paper is given to the spatially integrated radio emission of a single SNR and its temporal evolution, but also to the development of a consistent model that can be applied to any Type Ia SNRs.

Throughout our paper, we will use \lq classical\rq~non-linear diffusive shock acceleration (NLDSA) that naturally predicts that the spectrum steepens at low energies and flattens at higher energies, as the compression ratio felt by a diffusing particle depends on particle's energy. Recent gamma-ray observations of Galactic SNRs seriously challenge this understanding of CR acceleration at fast shocks.
There are evidences of high-energy part of CR spectra steeper than $E^{-2}$ (significantly steeper than what is expected in NLDSA that is implemented in our model) coming mainly from gamma-ray observations of SNRs \citep{caprioli12}. He developed a self-consistent scenario in which MFA induces the conditions for reversing previously mentioned trend and can lead to a steepening in the high energy part of the spectrum of CRs. The crucial point in \citet{caprioli12} is that he takes the Alfv\'en speed in the amplified magnetic field (as being $\delta B \gg B_0$) instead of calculation in the ambient field $B_0$. This is not crucial for this study and not modelled here because the part of the electron spectrum that matters for radio emission is at much lower energies. As for higher energies, it should be included in our future modelling, especially if we are interested in gamma-ray emission produced in hadronic scenario, with spectrum and emissivity directly related to the spectrum of highest energy CRs. The simplified model of gamma-ray evolution, used in our paper, is not strongly influenced by the shape of the spectrum as it approximates it with $E^{-2}$.

\subsection{Hydrodynamic modelling}
\label{sec:hydro} 
The dynamical evolution of an SNR was modelled by numerically solving the time-dependent HD equations of mass, momentum and energy conservation:

\begin{equation}
 \frac{\upartial}{\upartial t} \left(\!\!
  \begin{array}{c}
  \rho \\ \rho \bm{\upsilon} \\ E 
  \end{array}
 \!\!\right) + \nabla \cdot \left(\!\!\!
  \begin{array}{c}
  \rho \bm{\upsilon} \\
  \rho \bm{\upsilon} \bm{\upsilon} + \bm{I}P \\
  (E+P)\bm{\upsilon} \\
 \end{array} 
 \!\!\!\right)^\mathrm{T} \!\! = \left(\!\!
  \begin{array}{c}
   0 \\ 0 \\ 0
 \end{array}
 \!\!\right)
\label{eq2}
\end{equation}

\noindent where $\rho$ is the mass density, $\bm{\upsilon}$ is the flow velocity, $P$ is the thermal pressure, $\bm{I}$ is the unit vector, $E$ is the total energy density and $\gamma$ is the adiabatic index. The total energy density is the sum of the thermal and kinetic components:

\begin{equation}
 E=\frac{P}{\gamma-1}+\frac{1}{2}\rho \upsilon^{2}.
\label{baseeq}
\end{equation}

We adopt the {\scriptsize PLUTO} code \citep[Version 4.2;][]{mignone07, mignone12} to solve the system of HD conservation laws by using a cell-centred finite-volume approach based on Godunov-type schemes. The code design enables efficient usage of massively parallel computers
through the message passing interface standard (MPI) for interprocessor communications.

We do not include radiative cooling and thermal conduction in our HD equations, describing only the free and Sedov expansion phases of the SNR evolution in a tenuous, collisionless medium. The transition time from Sedov to radiative phase for an SNR is
described with the approximation
\citep[e.g.,][]{blondin98, petruk05, orlando11}

\begin{equation}
t\rs{tr} = 2.84\times 10^4\;E\rs{51}^{4/17}\;n\rs{H}^{-9/17}~\mbox{yr}~,
\label{trans_time}
\end{equation}

\noindent
where $E\rs{51}=E\rs{0}/(10^{51}\ \mbox{erg})$ and $E\rs{0}$ is the initial total explosion energy, contained mostly in form of kinetic energy. In our set of simulations, $t\rs{tr} > 0.5$ Myr and therefore our modelled SNRs never reach the radiative phase. We will compute the radio emission from the 
SNR but with assumption that the radiation has no impact yet on its
dynamical evolution.

Throughout our modelling, we do not activate the MHD solver of {\scriptsize PLUTO}
because it is in any case powerless in describing the generation of magnetic turbulence by CRs upstream of the shock and corresponding MFA. Such an amplified magnetic field is dominant compared to the field compressed only due to fluid compression, especially for young SNRs where non-linearity is
very pronounced. NLDSA module, which runs parallel with {\scriptsize PLUTO} HD code, simultaneously performs calculations of MFA, synthesize the global radio emission in this amplified field and also accounts for its impact on hydrodynamics.

We used the following set of {\scriptsize PLUTO} algorithms in our simulations: {\sf linear} interpolation with default limiter, {\sf HLLC} Riemann solver and {\sf RK2} algorithm for the time evolution.
Additionally, we used {\sf MULTID} shock flattening algorithm for the numerical dissipation near the strong shocks.
Our three-dimensional (3D) computations were carried out in spherical coordinates $(r, \theta, \phi)$, by using a static logarithmic grid, with mesh size increasing with the SNR radius.

Detection and tracking of the SNR shock waves in the fluid, travelling in some direction $x$, is based on two standard numerical conditions, namely $\nabla \bm{\upsilon} < 0$ and $\Delta x \frac{\nabla P}{P} > \epsilon_{\rm p}$, where $\epsilon_{\rm p}$ represents parameter, setting the shock strength.

We modified {\scriptsize PLUTO} modules\footnote{By default, algorithms in the {\scriptsize PLUTO} code are explicitly based on the assumption of a constant gamma law.} in order to couple the hydrodynamical evolution of the remnant with particle acceleration. Instead of 
a constant adiabatic gas index $\gamma$ (ratio of specific heats), obeying the ideal gas low $P = (\gamma-1) \bepsilon$, where $\bepsilon$ represents
thermal energy density, we adopted hydrodynamic equations to use the space and time-dependent adiabatic index $\gamma_{\rm eff} = \gamma_{\rm eff}(x,y,z,t)$ i. e. $P = (\gamma_{\rm eff}-1) \bepsilon$.
The effective adiabatic index $\gamma_{\rm{eff}}$, defined so it produces the same total compression $R_{\rm{tot}}$ as obtained from a non-linear model (described later in Section~\ref{sec:dsa}), is calculated at the shock front and then advected within the remnant, remaining constant in each fluid element as in \cite{ellison04} (see also \citealt{ferrand10,orlando12}). As pointed out
by \citet{ferrand10} and \citet{ferrand12}, each fluid element should remember the effect of shock modification induced by accelerated CRs at the time when 
shock wave passes through the fluid element. To fulfil this requirement, we threat the gas adiabatic index as a {\scriptsize PLUTO} built-in code feature called \lq passive scalar\rq~or \lq colour\rq~(denoted by $Q_{k}$) obeying the simple advection equation of the form

\begin{equation}
\frac{{\rm{D}} Q_{k}}{{\rm{D}} t} = 0,
\label{postadiab:tracer}
\end{equation}

\noindent which is added to the standard set of hydrodynamic equations (Equation~\ref{eq2}), where $\frac{\rm{D}}{\rm{D} t} = \frac{\upartial}{\upartial t} + \bm{\upsilon} \cdot \nabla$ denotes the Lagrangian time derivative.

The shock precursor is not explicitly modeled in the hydrodynamic
part of our simulations. The precursor properties are handled in a separate module containing a non-linear acceleration calculation. Non-linear effects on SNR hydrodynamics are visible only through the effective adiabatic index, as explained in the next section.

\subsection{Diffusive shock acceleration}
\label{sec:dsa} 

We model the evolution of an SNR, including the effect of the high-energy CR particles accelerated by the shock wave. It is widely accepted that the most efficient process of particle acceleration in SNRs is the diffuse shock acceleration (DSA), proposed by \citet{bell78a, bell78b} and \citet{blandford78}. Also known as a first-order Fermi mechanism\footnote{Enrico Fermi's idea that particles gain energy in collisions
with the moving irregularities of the magnetic field \citep{fermi49}, provides the basis of modern acceleration theories, including DSA.}, it provides the energy gain due to collisions with irregularities of the magnetic field of $\Delta E/E \propto u/\upsilon$, that is, first order in $u/\upsilon$, where $u$ is the velocity of magnetic perturbation and $\upsilon$ is the velocity of high-energy particle. For additional reviews of particle acceleration theories, see for example \citet{reynolds08, reynolds11} and \citet{urosevic14}.

In order to take into account the non-linear back reaction
of accelerated particles on the fluid structure, we use
the semi-analytical model of \citet{blasi02,blasi04} and
\citet{blasi05}. This model iteratively solves the particle 
distribution function $f(p)$\footnote{By definition, $f(p)$ satisfies $\mathrm{d} N = 4 \pi p^2 f(p) \mathrm{d} p$, where $N$ represents number of particles per unit volume. The energy distribution $f(E)$ can be calculated as $f(E)=4 \pi p^2 f(p)\frac{\mathrm{d} p}{\mathrm{d} E}$, giving $f(E) \propto p^2 f(p)$ in relativistic regime where $E \propto p$.} and the dimensionless fluid velocity $U(p)$, both as functions of particle momentum $p$. The boundary condition $U(p_{\rm max})=1$ has to be fulfilled since at $p > p_{\rm max}$ there are no CRs to contribute any pressure. The function $U(p)$ represents quantity defined as $U(p)=u_p/u_0$, where $u_p$ represents the average fluid velocity experienced by particles with momentum $p$ while diffusing upstream away from the shock surface and $u_0$ is the fluid velocity far upstream (shock wave velocity).

In the following, we will use standard indexing of quantities around
shock wave, namely subscript 1 (2) for parameters immediately upstream
(downstream), while subscript 0 denotes undisturbed, far upstream
quantities. We introduce three quantities $R_{\rm{sub}} = u_1/u_2$,
 $R_{\rm{tot}} = u_0/u_2$ and $R_{\rm{pre}}=u_0/u_1$ that are respectively the compression factor at the gas subshock, the total compression factor and compression in the shock precursor. The compression ratio $R_{\rm{sub}}$ ($R_{\rm{tot}}$) is expected to be lower (higher) than for the standard test-particle (TP) case where $R=4$. For a strongly modified shock, $R_{\rm tot}$ can attain values much larger than $R_{\rm sub}$.

The model iteratively solves acceleration, by numerical integration 
of integro-differential equations, providing the values
for shock compressions $R_{\rm sub}$, $R_{\rm pre}$ and $R_{\rm tot}$ that give $U(p)$ solution, satisfying $U(p_{\rm max})=1$.
This model only finds quasi-stationary
solutions so that we have to rerun it after each hydro time step 
(see recent paper \citealt{petruk16} that describes the time-dependent DSA at the non-relativistic shocks). Having the overall compression ratio, $R_{\rm tot}$, the effective ratio of specific heats is calculated as \citep{ellison04}

\begin{equation}
\gamma_{\rm eff} = \frac{M_{S,0}^2 (R_{\rm tot}+1) - 2R_{\rm tot}}{M_{S,0}^2(R_{\rm tot}-1)},
\label{postadiab:ellison04}
\end{equation}

\noindent where $M_{S,0}$ represents the sonic Mach number far upstream.
 
The originally proposed non-linear acceleration model of Blasi usually gives high levels of shock modification, as the total shock compression factors may exceed $\sim$ 50--100 \citep{amato05} and thus do not compare well with some observations, suggesting $R_{\rm{tot}} \sim$ 7--10 or slightly higher \citep[see e.g.,][]{volk05}. We assume that part of the energy in the form of turbulent Alf\'en waves, excited by energetic particles and responsible for the scattering of charged particles, is damped on the thermal gas and heats the gas in the upstream region. Our model uses non-adiabatic compression in the precursor proposed by \citet{berezhko99}

\begin{equation}
 \frac{P_p}{P_0} = U_p^{-\gamma} (1 + \zeta(\gamma -1)\frac{M_{S,0}^2}{M_{A,0}}(1 - U_p^{\gamma})),
\label{postadiab:nonadiab}
\end{equation}

\noindent which is caused by the Alvf\'en heating and significantly reduces the shock modification. Here, $P_0$ represents far upstream
fluid pressure, $P_p$ and $U_p$ are respectively the fluid pressure
and dimensionless fluid velocity at any point $x_p$ (for a given
diffusion law $D(p)$, particles of momentum $p$ diffuse up to a 
distance $x_p \sim D(p)/u_p$ where the fluid velocity is $u_p$) and 
$\zeta \in [0,1]$ is a free parameter introduced by \citet{caprioli09}.
This parameter accounts that the fraction $\zeta$ of the energy transferred from CR streaming
to MHD waves is dissipated as heat in the plasma by non-linear damping processes. 
The damping of the waves is mitigated for $\zeta < 1$  and therefore allows MFA\footnote{Otherwise, if we have $\zeta \simeq 1$, the rate of damping of the waves is close to that of wave-growth and the MFA is heavily suppressed.}.

We use a recipe for the injection of particles from the thermal pool, proposed by \citet{blasi05}, and set the following fraction $\eta$ of particles entering the acceleration process from Maxwellian downstream thermal pool

\begin{equation}
\label{eq:efficiency}
\eta = \frac{4}{3\sqrt{\rm \pi}} (R_{\rm sub}-1) \xi^3 e^{-\xi^2},
\end{equation}
\noindent which assumes that only particles with momentum $p_{\rm inj} \ge (\xi - u_2/c) \ p_{\rm th,2}$ can be involved in acceleration process, where $p_{\rm th,2} = \sqrt{2 m_{\rm p} k T_2}$ represents the mean downstream thermal momentum, $m_{\rm p}$ is the proton mass and $T_2$ is the downstream temperature. The shift $u_2/c$ is due to the assumption that thermal particles in downstream have a Maxwellian spectrum in the fluid reference
frame, while $p_{\rm inj}$ is taken in the shock frame.
The injection parameter $\xi$ strongly affects the acceleration fraction $\eta$. \citet{caprioli14a} conclude from their kinetic simulations that for parallel non-relativistic shocks a fraction of about $10^{-3}$ to $10^{-4}$ of the particles crossing the shock is injected into the DSA process and that the injection parameter is $\xi \simeq$ 3--4.

While the thermal leakage may represent a viable way of parameterizing injection, it does not account for the dependence of ion injection on shock inclination, elaborated by \citet{bell11}. Recent PIC simulations bring back again the dependence of ion injection on shock inclination \citep{CapPop15}. They show that ions are injected not by being heated and then leaking, but instead by specular reflection. A different dependence of $\eta$ on subshock compression is then hard to encompass in a formula similar to Equation~\ref{eq:efficiency}. This refinement is well beyond the goal of this paper, but should be somehow encompassed in future modelling.

It is worth stressing that using Equation~\ref{eq:efficiency} introduces some kind of time dependence of the acceleration efficiency, as compression at the subshock is not constant during SNR evolution. This may be seen as an improvement in comparison with previous models setting $\eta = \rm{const}$, keeping in mind that it may be still artificial and a questionable simplification. Recently proposed, a new theoretical model of time-dependent shock acceleration of particles \citep{petruk16} shows that variable injection could be a crucial element in explaining the X-ray and gamma-ray spectra of young SNRs, but not so important for radio spectra. 
Nevertheless, the time dependence in injection efficiency is hard to model, and even more how the proton-to-electron ratio (introduced later in Section~\ref{sec:radio}) depends on time. Kinetic simulations, 
providing us first-principles calculations, seem to show that
ion injection does not vary as long as the shock is strong, while the
electron injection efficiency in the regimes considered here is still questionable.

We assume that the CR distribution vanishes at a distance $\sim \chi_{\rm{esc}} R_{\rm{s}}$ upstream of the shock wave, where parameter $\chi_{\rm{esc}}<1$ is the fraction of the shell radius $R_{\rm{s}}$, to account for the presence of a free-escape boundary beyond which highest energy CRs, mainly consisting of protons, cannot diffuse back at the shock and escape into the interstellar medium (ISM) \citep{caprioli10, morlino12}. This approximation allows us to determine the maximum momentum $p_{\rm{p,max}}$ of accelerated protons by assuming

\begin{equation}
\frac{D(p_{\rm{p,max}})}{u_0} = \chi_{\rm{esc}} R_{\rm{s}},
\label{postadiab:pmax}
\end{equation}

\noindent where $D(p)$ is the Bohm-like diffusion constant i.e. $D(p) = \frac{1}{3} \upsilon(p) r_{\rm{L}}(p)$, with $\upsilon(p)$ and $r_{\rm{L}}(p)$ are respectively the particle velocity and the Larmor radius, in agreement with the approach of \citet{bell13} and references therein. 
We use $\chi_{\rm{esc}} = 0.1$ in our model as suggested by \citet{morlino12}, which satisfies the condition that the acceleration time up to $p_{\rm{p,max}}$ is less than the age of the system \citep{blasi07}.

\subsection{Magnetic field amplification}
\label{sec:mfa} 

Galactic CR acceleration to the knee in the spectrum at a few PeV is only possible if the magnetic field ahead of an SNR shock is strongly amplified by CRs escaping the SNR \citep{bell13}.
Consistent calculation of the magnetic field strength, although not yet fully understood, is of the utmost importance for the radio emission of energetic electrons. Due to this, we include MFA in our model.

The MFA is driven by streaming instabilities, induced by CRs in the vicinity of SNR shocks. Instabilities can be resonant \citep{bell78a}, assuming that Alfv\'en waves, generated by particles streaming faster than the Alfv\'en speed, have a wavelength in resonance with the CR Larmor radius, and strongly driven, nearly purely growing, non-resonant \citep{bell04}. Non-resonant instabilities are not accurately described as Alfv\'en waves and they grow preferentially at wavelengths that are not resonant with the CR Larmor radius. \citet{amato_blasi09} showed that the
non-resonant modes are bound to be relevant mostly in the earlier 
stages of the SNR evolution, namely free expansion and early Sedov--Taylor, while streaming instability should be dominated by resonant waves for the most of the history of the SNR. However, according to \citet{caprioli14b}, their equation obtained for resonant instability fits well up to simulations with $M_A =100$ and can also be extrapolated to higher Mach numbers, inferred at the blast waves of young SNRs, in the case of efficient CR acceleration, which is certainly the case for G1.9+0.3. For such high-$M_{\rm{A}}$ shocks, according to \citet{caprioli14b}, we can distinguish two regions: the far upstream region dominated by non-resonant instability, and the precursor, where resonant and non-resonant instabilities grow at a comparable rate. Closer to the SNR shock, resonant instabilities seems to take over but ambient magnetic field is already considerably modified in the precursor. It is, however, hard to simulate what happens for $M_A \gg 100$ in global shock simulations mentioned, as simulations become quite expensive, and many questions remain open here. Due to this, whichever model we choose to implement, it will contain a lot of uncertainties until the future PIC simulations and better theory give us and improved understanding. 

We choose to model MFA due to resonant streaming instability,
being already compatible with Blasi's formalism and easy to implement in our code. With the assumption that all the turbulence is generated via streaming instability in the precursor, it is described 
by \citet{caprioli09} in the form

\begin{equation}
\label{eq:Pw-res}
\frac{P_{\mathrm{w},p}}{\rho_{0}u_{0}^{2}}=\frac{1-\zeta}{4M_{A,0}} U_p^{-3/2}(1-U_p^{2}),
\end{equation}

\noindent This model of MFA is also used by \citet{lee2012} and
\citet{ferrand14}.  Here, $P_{\mathrm{w}} =\frac{1}{8\pi} (\sum_{\mu}^{}\delta B_{\mu})^2 $ denotes the precursor magnetic pressure of 
Alfv\'en waves (subscript $\mu$ indicates modes of the magnetic turbulence) and $M_{A,0}=u_0/\upsilon_{\rm A}=u_0 \sqrt{4 \pi \rho_0}/B_0$
is the Alfv\'enic Mach number far upstream. For simplicity, we do not include pre-existing magnetic turbulence in the ISM in our models. The factor $(1-\zeta)$
is introduced to balance the factor $\zeta$ in \autoref{postadiab:nonadiab}, and accounts for local wave dissipation and reduction of MFA.
The total magnetic field at point $x_p$ is then
calculated with

\begin{equation}
\label{eq:totalB}
B_p^2 = B_0^2 + 8\pi P_{\mathrm{w},p},
\end{equation}

\noindent with $B_0$ denoting the ordered component of the ambient
magnetic field. However, keep in mind that this is a rough estimate as the effective ambient $B_0$ field that we need in the precursor is always determined by the Bell's non-resonant instabilities far upstream and will be larger than the few $\mu$G of the Galactic field.

The magnetic pressure in the amplified fields becomes quite high, comparable with or even higher than the thermal one, making the dynamical role of amplified magnetic fields non-negligible \citep{caprioli09}.
Due to this, the magnetic pressure as well as the pressure of accelerated particles is accounted for in the NLDSA part of the presented calculations
and then affects fluid compressibility through the effective adiabatic index. The global shock modification follows from the conservation of momentum between the far upstream medium and any precursor point $x_p$,
which involves four terms: dynamical pressure $\rho u^2$, thermal pressure 
$P$, non-thermal pressure of CRs computed from their distribution $f(p)$ as
$P_{\rm{cr}}=\frac{4\pi}{3}\int{p^2 f(p) p \upsilon(p) \mathrm{d} p}$ and magnetic pressure $P_{\rm{w}}$ 
\citep{ferrand14}, described with equation

\begin{equation} 
\label{eq:conservation}
\rho_p u_p^2 + P_{\mathrm{th},p} + P_{\mathrm{cr},p} + P_{\mathrm{w},p} = \rho_0 u_0^2 + P_{\mathrm{th},0} + P_{\mathrm{cr},0} + P_{\mathrm{w},0}.
\end{equation}

\noindent As we will see later, the factor $\zeta$ in our simulations is close to $1/2$ and therefore gives the amplified magnetic field an important role in SNR dynamics.

Also note that MFA leads to non-adiabatic heating of the background plasma (turbulent heating, as mentioned in Section~\ref{sec:dsa}) in such a way that plasma and magnetic field pressure are almost in equipartition throughout the precursor \citep{caprioli14a}.

The maximum value of the amplified upstream magnetic field $B_1$ is 
reached immediately ahead of the shock wave and we calculate it 
by putting $U_{p} = u_1/u_0 = 1/R_{\rm pre}$ in \autoref{eq:Pw-res}. We
use a common assumption that the $B$-field is totally random in orientation, 
as a consequence of the strong turbulence.
 We then assume that the downstream magnetic field is compressed only due to fluid compression such that the components in a shock plane are compressed and the three components of the magnetic field are roughly equal. The downstream magnetic field is then given by $B_2=B_1 \sqrt{1/3 + 2/3 R_{\rm sub}^2}$.

We also account for damping of the amplified magnetic field in
the downstream region. We use the following recipe for the downstream 
magnetic field \citep{morlino12}, based on the non-linear
Landau damping mechanism

\begin{equation} \label{eq:landau}
B_d(r) \simeq B_2 \exp\left(-\frac{Rs-r}{\lambda_{\rm{nl}}}\right).
\end{equation}

\noindent The typical length scale $\lambda_{\rm{nl}}$ for the
non-linear Landau damping given by

\begin{equation} \label{eq:landau-scale}
\lambda_{\rm{nl}} = \frac{3 \chi_{\rm esc}}{0.05} \frac{u_0^2}{\upsilon_A(B_2) c} R_s,
\end{equation}

\noindent where $\upsilon_A(B_2)$ is the Alfv\'en velocity in the downstream region.

Bell's model for the MFA due to the non-resonant streaming instability \citep{bell04} predicts the total saturated magnetic energy density 

\begin{equation} 
\label{eq:bell}
\frac{B_{\rm{sat}}^2}{2\mu_0} \sim \frac{1}{2} \frac{u_0}{c} \bepsilon_{\rm{cr}},
\end{equation}

\noindent where $\bepsilon_{\rm{cr}}$ represent the CR energy density at the shock. In the active SNR phase, when $\bepsilon_{\rm{cr}} \sim \rho_{0}u_{0}^{2}$, this results in an amplified magnetic field of
$B \propto u_{0}^{3/2}$. Another point of view, namely equipartition between the total energy densities of CRs and that of the magnetic field \citep{bik05, arbo12} results in $B \propto u_{0}$.
Both dependences, mentioned above, were implemented in the numerical model of \citet{ksenofontov10}, while \citet{bif04} used only the later one. 
\citet{vink12} favors a dependence $B \propto u_{0}^{3/2}$, providing the observational evidence, but also notes that the dynamic range makes the dependence on shock velocity uncertain. On the other hand, $B \propto u_{0}^{3/2}$ implementation in \citet{ksenofontov10}, results in a too slow increase in radio flux density of SNR G1.9+0.3 to provide a reasonable agreement with observations.

Equation~\ref{eq:Pw-res}, used as a receipt for MFA in our modelling of radio evolution, is a pretty much non-linear and therefore the magnetic field dependence on shock velocity will be affected by other simulation
parameters. If we deduce from Equation~\ref{eq:Pw-res} that $P_{\mathrm{w},p} \sim \frac{\rho_{0}u_{0}^{2}}{M_{A,0}}$, we obtain 
$B \propto (u_0 B_0)^{1/2}$ that should be taken with caution as the term 
$U_p^{-3/2}(1-U_p^{2})$ probably brings additional dependence on shock velocity. We will see later in Section~\ref{sec:results} the relation between the amplified magnetic field and the shock velocity, obtained from our simulations.

The previous paragraph demonstrates an important difference between 
MFA driven by resonant and non-resonant streaming instabilities. 
The saturation of the resonant CR-driven instability explicitly depends on
initial $B_0$, which is not the case for Bell's non-resonant 
instabilities \citep[see, ,e.g. equations 12 and 13 in][]{amato11}.

Magnetic field modelling here does not consider its stretching and amplifying caused by the initial clumping of the ejecta and due to the development of Rayleigh--Taylor instabilities \citep[see e.g.][]{orlando12}. It is thus important to simulate these effects by using {\scriptsize PLUTO} MHD in future, in order to follow morphological evolution of SNR along with the integrated radio emission that is modelled here.


\subsection{Radio emission}
\label{sec:radio} 

Although the injection mechanisms for electrons are much 
less clear than for protons, we use the assumption that electron injection is the same as that of protons and normalize their spectrum with respect to the protons' spectrum:

\begin{equation} \label{eq:common}
f_{\rm e} (p) = K_{\rm ep } f_{\rm p} (p),
\end{equation}

\noindent where parameter $K_{\rm ep}$ represents the electron-to-proton
ratio, very likely related to the different mechanisms responsible for lepton and hadron injection. We allow this parameter to vary from the value $K_{\rm ep} = 10^{-2}$, observed near Earth in the diffuse spectrum of Galactic CRs around GeV energies. Since the energy losses of GeV electrons are not significant for propagation in the Galaxy, the electron-to-proton ratio is also about 10$^{-2}$ in the source. Nevertheless, this does not mean that this ratio must be the same for young SNRs \citep{zirak08}. Young SNRs, like G1.9+0.3, may be the main source of Galactic electrons with energies higher than 10 TeV, while GeV electrons may be produced in older SNRs and therefore the value measured in CRs seems to be determined by later stages of the SNR evolution \citep[e.g.,][]{sarb17}. From recent PIC simulations of simultaneous acceleration of protons and electrons \citep{park15}, the CR electron-to-proton ratio is
inferred to be $K_{\rm{ep}} \approx 10^{-3}$ to $10^{-2}$, for 
shock velocity $u_0/c \approx$ 0.02--0.1 and reduced electron to proton mass $m_{\rm p}/m_{\rm e}$ ranging from 100 to 400\footnote{Authors find marginal change of $K_{\rm{ep}}$ when increasing $m_{\rm p}/m_{\rm e}$ from 100 to 400}. We assume that the spectrum of accelerated electrons is parallel to protons' one, except for large momenta, since the DSA mechanism should not be dependent on charge. We neglect the dynamical role of electrons in our model.

Assumption about parallel proton and electron spectra holds as long as synchrotron losses
are neglected. However, at energies around TeV, electrons suffer
synchrotron losses that can be consistently taken into account by supplementing the
ordinary diffusive transport equation by a corresponding loss term \citep[as done, e.g. in][]{bif02,bif04}. Strict numerical treatment of electron cooling is beyond the scope of this paper, as for radio-emitting electrons it can be safely neglected. Nevertheless,
we implement the simple \lq toy\rq~model, dealing with high-energy part of the electron spectrum, in
order to obtain satisfying model of synchrotron spectra from radio to X-ray domain, which is later demonstrated
in Section 3. We assume that electron spectra above a certain energy becomes steeper,
i.e. changes from $p^{-q}$ to $p^{-(q+\delta)}$ \citep{tanaka08,longair11} and allow $\delta$ to be different than 1. Following \citet{tanaka08}, we calculate the position of the transition from an uncooled to a cooled
regime as

\begin{equation} \label{eq:Eb}
E_{b} = 1.25 \left(\frac{B_2}{100 \ \mu \rm{G}}\right)^{-2} \left(\frac{t_{0}}{10^3 \rm{yr}}\right)^{-1} \ \rm{TeV},
\end{equation}

\noindent where $t_{0}$ represents the current SNR age. Going from $E_b$ (corresponding momentum $p_{b}$) to higher energies, we steepen previously calculated electron spectra and apply the following form of cut-off $\exp\left[-(p/p_{\rm{e,max}})^2\right]$, as suggested by \citet{zirak07} for the loss-dominated case.

We calculate the electron maximum momentum $p_{\rm{e,max}}$, in the Bohm diffusion regime, by using the approximate implicit expression, determined by \citet{morlino09}. This
approach is based on equating the acceleration time with the minimum between the time for energy losses and the age of the SNR, when only synchrotron losses are important,

\begin{equation} \label{eq:pmax_e_2}
 p_{\rm e,max}= \frac{3}{2} \sqrt{\frac{m_e^3 c^4}{e B_1 r_0}}\,
 \frac{u_0}{c}\, U_p(p_{\rm e,max})\, \sqrt{\frac{1-R_{\rm tot}^{-1}\,U_p^{-1}(p_{\rm e,max})}{1+R_B R_{\rm tot} U_p(p_{\rm e,max})}} \,,
\end{equation}
\noindent where $r_0$ is the classical electron radius, $R_B$ is the magnetic field compression factor at the  sub-shock.

The spectrum for electrons at the shock is then 
\begin{equation} 
\label{eq:f_e(p)}
 f_{\rm e,0}(p)= K_{\rm ep}\,f_{\rm p,0}(p) p^{-\delta} e^{-p^2/p_{\rm e,max}^2} ,
\end{equation} 

\noindent where $\delta=0$ for momenta $p<p_b$ and $\delta>0$ for 
$p \ge p_b$. We do not expect a sharp break in the energy spectra for the electrons, but rather some steepening \citep{blasi10}. Due to this, for our \lq toy\rq~model, $\delta$ has to be a continuous function of momenta, which is later obtained as the best fitting to the observed spectra. The standard assumption has been applied that the spatial distribution of particles for a plane shock is constant downstream and drops exponentially in the upstream \citep{reynolds08}. Adiabatic losses are not taken into account as we assume that most of the radio emission comes from the relatively thin shell near FS.

The total volume emissivity (power per unit frequency interval per unit
volume) of the relativistic electrons is defined as

\noindent
\begin{equation} \label{eq:emiss1}
\epsilon_{\nu} = \int_{E} P(\nu) N(E) \mathrm{d} E.
\end{equation} 

\noindent Here, $P(\nu)$ is the total emissivity of a single electron of energy $E$ which has a pitch angle $\vartheta$ with respect to the magnetic field given by \citep{wilson13}

\noindent
\begin{equation} \label{eq:emiss2}
P(\nu) = \frac{\sqrt{3} e^3 B \sin{\vartheta}}{m_e c^2} F\left(\frac{\nu}{\nu_c}\right),
\end{equation} 

\noindent where $\nu_c=\frac{3eB_{\perp}p^2}{4\pi (m_e c)^3}$ is the electron critical frequency and $F(x)$ is a synchrotron function defined as
 
\begin{equation} \label{eq:sychr}
F(x) = x\int_{x}^{\infty} K_{5/3}(x') \mathrm{d} x',
\end{equation}

\noindent with $K_{5/3}(x)$ being the modified, non-integer order Bessel
function \citep{abram72}. With an accuracy of less than 0.6 per cent, the synchrotron function $F(x)$ can be approximated with a linear combination of its known approximations for $x \ll 1$ and $x \gg 1$, derived by \citet{fouka13}.

Hence, by combining these relations, the SNR total luminosity at frequency 
$\nu$ is calculated from the obtained electron spectrum $f_{\rm{e,0}}(p)$ by using the following expression 

\begin{equation} \label{eq:luminosity}
L_{\nu} = \frac{16 \pi^2 \sqrt{3} e^3}{m_e c^2} \int_{R_{\rm cd}}^{R_s} B_{\perp} r^2 \mathrm{d} r 
\int_{p_{\rm inj}}^{p_{\rm e,max}} p^2 f_{\rm{e,0}}(p) F\left(\frac{\nu}{\nu_c}\right) \mathrm{d} p,
\end{equation}

\noindent where $B_\perp$ is the magnetic field component perpendicular to the line of sight (LoS) and we shall use $B_\perp (r) = 0.5 B_d (r)$.

Our model uses a reasonable approximation that the radio emission from accelerated electrons comes only from the shocked ISM located between contact discontinuity  (at radius $R_{\rm cd}$) and forward shock (FS). This assumption is based on the fact that, because of the compression of magnetic field in downstream, the overall synchrotron radiation of the SNR is dominated by the emission originating from the downstream region.

In order to obtain the radio flux density $S_\nu$ at a given SNR distance $d$, we use the following relation:

\begin{equation} \label{eq:snu}
S_\nu = \frac{L_\nu}{4 \pi d^2}.
\end{equation}

Many authors have implemented an idea that CR acceleration may also occur at the reverse
shock (RS) of young SNRs \citep[and others]{ellison05, zirak10, zirak12} although 
it is hard to expect a very large magnetic field in the ejecta into which RS propagates. The magnetic field frozen in the ejecta dilutes orders of
magnitude below levels required to accelerate particles, during the early expansion phase of SNR.
\citet{casa01} have identified the FS and RS in Cassiopeia A and showed that
radio (together with Si) emissivity radial profile shows a sharp rise at what
they characterize as the RS. However, \citet{morlino12} concluded that there is no
evidence of DSA at the RS, for the particular case of relatively young Tycho SNR, by investigating its radial profile of the radio emission. Similar to the latter case, currently available VLA and ATCA radial profiles of G1.9+0.3 radio emission \citep{green08, horta14}, taken between 1984 and 2009, do not show any emission that could be linked with the RS. Therefore, in this paper we only account for the radio emission from FS of SNR G1.9+0.3 and it seems enough to account for the observed radio flux.

\subsection{Simple estimates of the gamma-ray emission}
\label{sec:gamma}

During the last decade, new generations of gamma-ray telescopes operating in the GeV and TeV range provided us new insights into SNR phenomenology and CR acceleration. It is generally accepted that two distinct physical 
mechanisms are responsible for gamma emission. Electrons produce a gamma radiation via inverse Compton (IC) scattering on different microwave, IR and optical photons, in the so-called leptonic scenario. The contribution of relativistic bremsstrahlung is also produced in leptonic scenario, but it is usually negligible in SNRs. In the second case, so-called hadronic scenario, gamma rays are produced by the decay of neutral pions ($\pi^0$) produced in collisions between CRs and the background gas.

Despite relatively deep exposures, the High Energy Stereoscopic System (H.E.S.S.) data did not show any signs of significant TeV gamma-ray emission from SNR G1.9+0.3 \citep{hess14}.

As pointed out by \citet{ksenofontov10}, the TeV gamma-ray flux is expected to increase with time as well as the radio flux, mainly due to the increase in the overall number of CRs with energy above 10 TeV. It will be interesting to estimate the gamma-ray luminosity of the SNR, as a function of time and whether it could be visible in TeV gamma-rays by future instruments like the Cherenkov Telescope Array (CTA).

Consistent calculation of broad-band gamma emission of G1.9+0.3 is far from the scope of this paper. Due to this, we will rather use the simplified model of \citet{zirak08} which assumes the $E^{-2}$ spectrum of highest energy protons at the shock front. This may not be so crude estimate, as \citet{caprioli12} noted that slope predicted by the standard NLDSA is even farther from the required $E^{-2.1}-E^{-2.2}$ to account for the observed gamma-ray phenomenology.

The differential gamma-ray flux from the pion decay at energies in the range $m_pc^2 < E < 0.1E_{\rm{p,max}}$ (where $E_{\rm{p,max}}$ represents the maximum proton energy) may be estimated as proposed by \citet{zirak08}:

\begin{equation}
E^2F_{pp}(E)=\frac {R_s^3K_{\rm{\pi\pi}}\sigma_{\rm{pp}}c n_{\rm{H}}^2\xi_{\rm{cr}}m_p u_0^2}{d^2\ln (p_{\rm{p,max}}/m_p c)}
\left( 1+4\frac {n_{\rm{He}}}{n_{\rm{H}}}\right) ^2.
\end{equation}
Here, $K_{\rm{\pi\pi}}=0.17$ is the fraction of the proton energy
transmitted to the parent neutral pions, $\sigma _{\rm{pp}}$ is the
total inelastic $p$--$p$ cross-section, $\xi _{\rm{cr}}$ is
the ratio of the CR pressure downstream of the shock $P_{\rm{cr}}$ to the dynamical pressure $\rho _0 u_0^2$, while $n_{\rm H}$ and $n_{\rm He}$ represent hydrogen and helium number density in ISM, respectively, assumed to be partitioned as $n_{\rm H}:n_{\rm He}=9:1$. We use the value $\sigma_{\rm{pp}}=37.4~\rm{mb}$ for our estimate of the 4 TeV gamma-rays flux \citep{kelner06}. It was assumed that the accelerated protons fill the SNR uniformly.

The differential flux of gamma-rays from the IC scattering in the synchrotron losses dominated case may be estimated as proposed by \citet{zirak08}:

\begin{equation}
\label{eq:Fic}
E^2F_{\rm{IC}}(E)=\frac {3\xi _{\rm{cr}}}{8\xi _B}\frac {K_{\rm{ep}}R_s^2U_{\rm{rad}}u_0}{d^2\ln (p_{\rm{p,max}}/m_p c)}.
\end{equation}
Here, $\xi _B$ is the ratio of the magnetic energy $B^2/8\pi $ to the dynamical pressure $\rho _0 u_0^2$ and $U_{\rm{rad}}$
is the energy density of the scattered photons, computed here only for cosmic microwave background (CMB) radiation $U_{\rm{rad}}=(4\sigma/c)T^4 \approx 4.2 \times 10^{-13}~\rm{erg}\,cm^{-2}s^{-1}$. The flux given in Equation~\ref{eq:Fic} is valid for energies smaller than the cut-off energy given by

\begin{equation}
E_c \approx 5~\rm{TeV} \left(\frac{u_0}{3000~km/s}\right)^2 \left(\frac{B_2}{100~\mu G}\right)^{-1} \frac{T}{2.7 K},
\end{equation}
where $T$ represents the temperature of the scattered photons. We checked a posteriori that the gamma-ray energy around few TeV satisfies this requirement up to around 2500 yr of simulated SNR evolution.

\section{Results and discussion}
\label{sec:results}

VLA observations from 1985 \citep{green04} show a strong asymmetry in the shell at 21 cm, perhaps indicative of an external density gradient. The mean radius of the bright X-ray ring is about 2 pc, but with the east and west \lq ears\rq~at about 2.2 pc \citep{reyn08}. We neglect any possible gradient of ambient density in surrounding ISM, which leads to a complicated morphology \citep[see][]{orlando07} and seek for a global qualitative description of integrated continuum radio emission.

Trying to model the observed radio morphology as a consequence of hypothetical global magnetic field gradient would be tricky due to the existing disagreement between observations and theory. Young SNRs have a predominantly radial magnetic field structure, visible through polarization measurements \citep{helder12,reynolds12}. On the other hand, modern theories and simulations of MFA predict a strong turbulence of amplified field \citep{bell04, caprioli14b} as a result of the interaction of CRs with the upstream plasma and ambient field.

We performed 3D HD simulations describing the expansion of the SNR 
G1.9+0.3 in spherical coordinates with the {\scriptsize PLUTO} code, adopted according 
to the model described in the previous section. As the magnetic field does not play a dynamical role in the evolution of the SNR, we do not use MHD modules existing in {\scriptsize PLUTO}. However, we calculate the magnetic field strength and its amplification by using our separate NLDSA modules tied to {\scriptsize PLUTO}, as it is necessary for CR acceleration as well as for the radio emission.

\begin{figure}
	\includegraphics[width=1.05 \columnwidth]{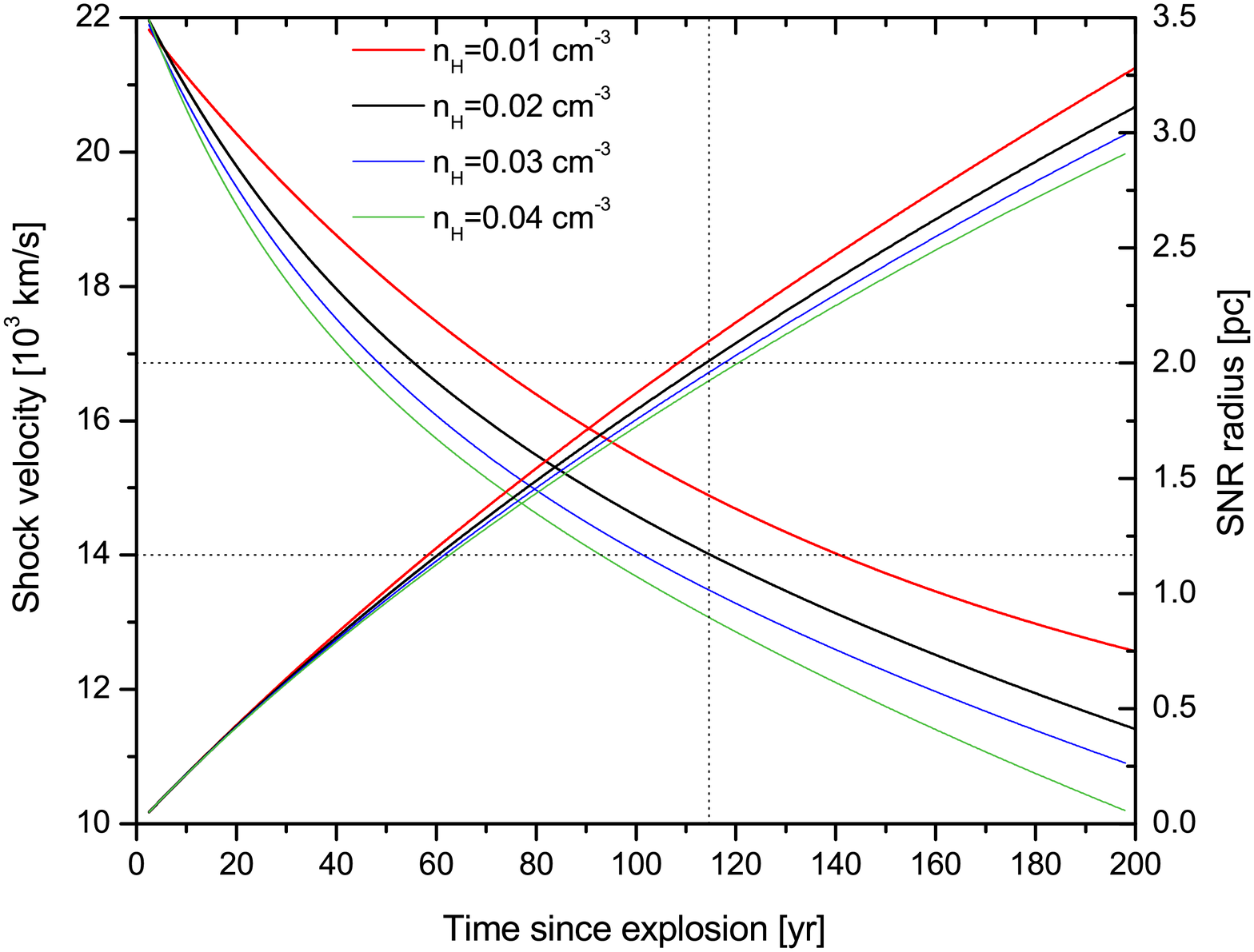}
    \caption{Evolution of a G1.9+0.3 shock velocity and radius, obtained from 3D HD simulations including efficient DSA. The left axis represents shock velocity in units of 10$^3$ km/s, which is a decreasing function of time. The right axis represents the radius of expanding SNR in parsecs. Radius and shock velocity evolution curves correspond to ISM hydrogen number densities of 0.01 (red thick line), 0.02 (black), 0.03 (blue) and 0.04 (green) cm$^{-3}$. Horizontal dotted lines correspond to currently available measurements of the mean radius and shock velocity from 2008. Vertical dotted line marks the epoch $t_{\rm{SNR}}=115$ yr when observed and simulated radius and shock velocity coincide, i.e. represent our inferred age of the SNR. The starting time and radius for the simulation are respectively $t_0 \approx 2.5$ yr and $R_0 \approx 0.05$ pc.}
\label{fig:hydro}
\end{figure}

Our initial conditions were chosen in order to reproduce G1.9+0.3
after around 100 yr of evolution\footnote{\citet{horta14} put the highest upper age limit of 180 yr, assuming a constant expansion rate since
the SN event.} in terms of shock radius, which is about 2 pc (near 100 arcsec in diameter) for an assumed location near the GC \citep{reynolds08}, and shock velocity of 14\,000 km s$^{-1}$, deduced mainly from Fe emission with a width of about 28\,000 km s$^{-1}$\citep{borkow10}.
In all of our simulations, the ambient magnetic field strength is set to
value $ B_0 = 5 \mu$G, representative of the average Galactic field. For the initial density structure of the ejecta, we used the exponential profile that has been shown to be the best approximate
representation of explosion models for Type Ia SNe \citep{dwarka98},
thought to represent thermonuclear disruption of a white dwarf.
We add clumps in the initial ejecta as per-cell random density perturbations \citep{orlando12} and they trigger Rayleigh--Taylor instability at the contact discontinuity. We assumed an initial spherical remnant with a radius of 0.05 pc (corresponding to an initial age of $\approx$ 2.5 yr), ejecta mass equal to the Chandrasekhar mass $M_{\rm ej} = 1.4 M_{\sun}$ and the total explosion energy $E_0=10^{51}$ erg. SNR expands through a homogeneous isothermal plasma with temperature $T = 10^4$ K (corresponding to an isothermal sound speed $c_{\rm S} = 9.9$ and Alv\'en speed $\upsilon_{\rm A}=39$ km/s\footnote{Velocity 14\,000 km/s therefore gives far upstream sonic Mach number $M_{\rm S} \approx 1410$ and Alfv\'enic number $M_A \approx 360$.} for ambient density 0.02 cm$^{-3}$) and mass density $\rho_0 = \mu m_{\rm{H}} n_{\rm{H}}$, characterized by the hydrogen number density $n_{\rm{H}}$, where $\mu=1.4$ is the mean atomic mass (assuming cosmic abundances) and $m_{\rm{H}}$ is the mass of the hydrogen atom.

\begin{figure}
	\includegraphics[width= \columnwidth]{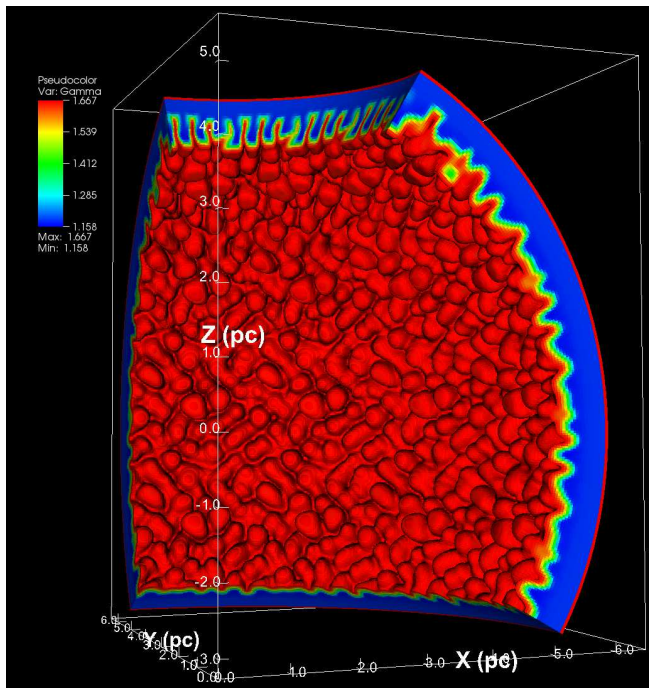}
    \caption{3D rendering of the spatial distribution of the plasma
     effective adiabatic index for a model accounting for the shock modification by accelerated CRs. Simulation describes the expansion of 
SNR through the ISM. Credit by VisIt.}
    \label{fig:3D-SNR}
\end{figure}

In order to estimate the ISM density and age of SNR, we performed a set of 3D HD simulations with different ambient hydrogen number densities ranging from 0.01 cm$^{-3}$ to 0.04 cm$^{-3}$ (Fig.~\ref{fig:hydro}). We simulate one octant of the remnant, for the total time of 200 yr, with a resolution of 2048 $\times$ 512 $\times$ 512 grid cells, respectively, for each of the spherical coordinates $r$, $\theta$ and $\phi$ (Fig.~\ref{fig:3D-SNR}). Soon after the explosion, SNR dynamical evolution is characterized by increasing the radius and decreasing the shock velocity\footnote{Note that the shock velocity is not the fluid speed of any material present in our simulation box, but an interface between different conditions that
is propagating through the fluid.}. The observed shock velocity of 14\,000 km/s and SNR radius of 2 pc have to be reached in our simulations at the same time after the explosion, which we take as the SNR age. The hydrogen number density satisfying this requirement is $n_{\rm H} =$ 0.02 cm$^{-3}$ and the corresponding SNR age is 115 yr, for the epoch 2008 when observations, we used for comparison with models, are made. This implies an explosion date of about 1893 and a current age of 123 yr.

As referent radio data for SNR G1.9+0.3 we use observations
made by \citet{green08} at 1.43 and 4.86 GHz with the VLA, which are respectively 0.935 $\pm$ 0.047 Jy and 0.437 $\pm$ 0.022 Jy. Combining these integrated flux densities gives a steep radio emission spectral index $\alpha$ (defined so that the flux density scales with frequency as $S_\nu \propto \nu^{-\alpha}$ ) of 0.62 $\pm$ 0.06, using the assumed 5 per cent statistical uncertainties in the individual flux densities. In general, observations confirm that young SNRs have radio spectral indices steeper than the expected $\alpha=0.5$ \citep{urosevic14}, derived from TP DSA \citep{bell78a,bell78b}. \citet{bell11} showed that young SNRs with the quasi-perpendicular orientation of the magnetic field should have
steeper spectral indices. For a detailed review on radio spectra of SNRs and some other possible explanations for steep spectra of young SNRs, see \citet{urosevic14} and references therein. Some properties of the
time-dependent solutions (instead of quasi-stationary solutions used in our modelling) could also be responsible for the deviation of the observed radio index from the classical value 0.5 in some young SNRs, as recently shown by \citet{petruk16}.

We also pose a radio light curve for G1.9+0.3, observed with the Molonglo Observatory Synthesis Telescope (MOST), spanning 20 yr from 1988 to 2007 at a frequency of 843 MHz \citep{murphy08}. Two most recent measurements (closest to the time when VLA observations have been made), 0.97 $\pm$ 0.11 Jy from epoch 2007.430 and 1.32 $\pm$ 0.09 Jy from epoch 2007.463, will only be used for comparison with our best-fitting modelled spectra made using VLA measurements \citep{green08}, although showing evident inconsistency and large measurement errors. Change in the radio flux should not be neglected for earlier MOST measurements as they cover around one-sixth of the estimated lifetime of the SNR.

\citet{horta14} also obtained radio
flux density measurements but significantly smaller ($\sim$ 50 per cent) than VLA measurements. They attribute this large difference to missing short spacings and poorer $uv$ coverage of the ATCA images.

\begin{figure}
	\includegraphics[width=0.9 \columnwidth]{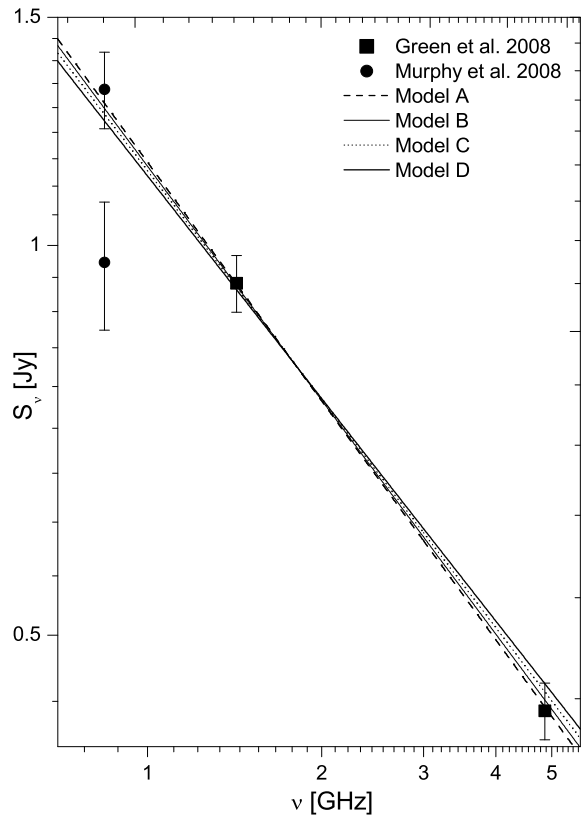}
    \caption{Measured radio fluxes at 843 MHz \citep[MOST]{murphy08} and 
    1425 and 4860 MHz \citep[VLA]{green08}; and four modelled 
    radio spectra, corresponding to models from Table~\ref{tab:models}.
    We include only two MOST fluxes at 843 MHz for comparison, 
    observed in epochs 2007.46 and 2007.43, closest to the VLA observing 
    epoch of 2008. Change in the radio flux should not be neglected for
     earlier MOST measurements.
    The ambient magnetic field is set to value 
    $B_0=5 \mu$G. Observed and modelled radio fluxes correspond
    to epoch 2008, when G1.9+0.3 was around 115 yr old, as obtained
    from our simulations.}
    \label{fig:fit}
\end{figure}

\begin{figure}
	\includegraphics[width=\columnwidth]{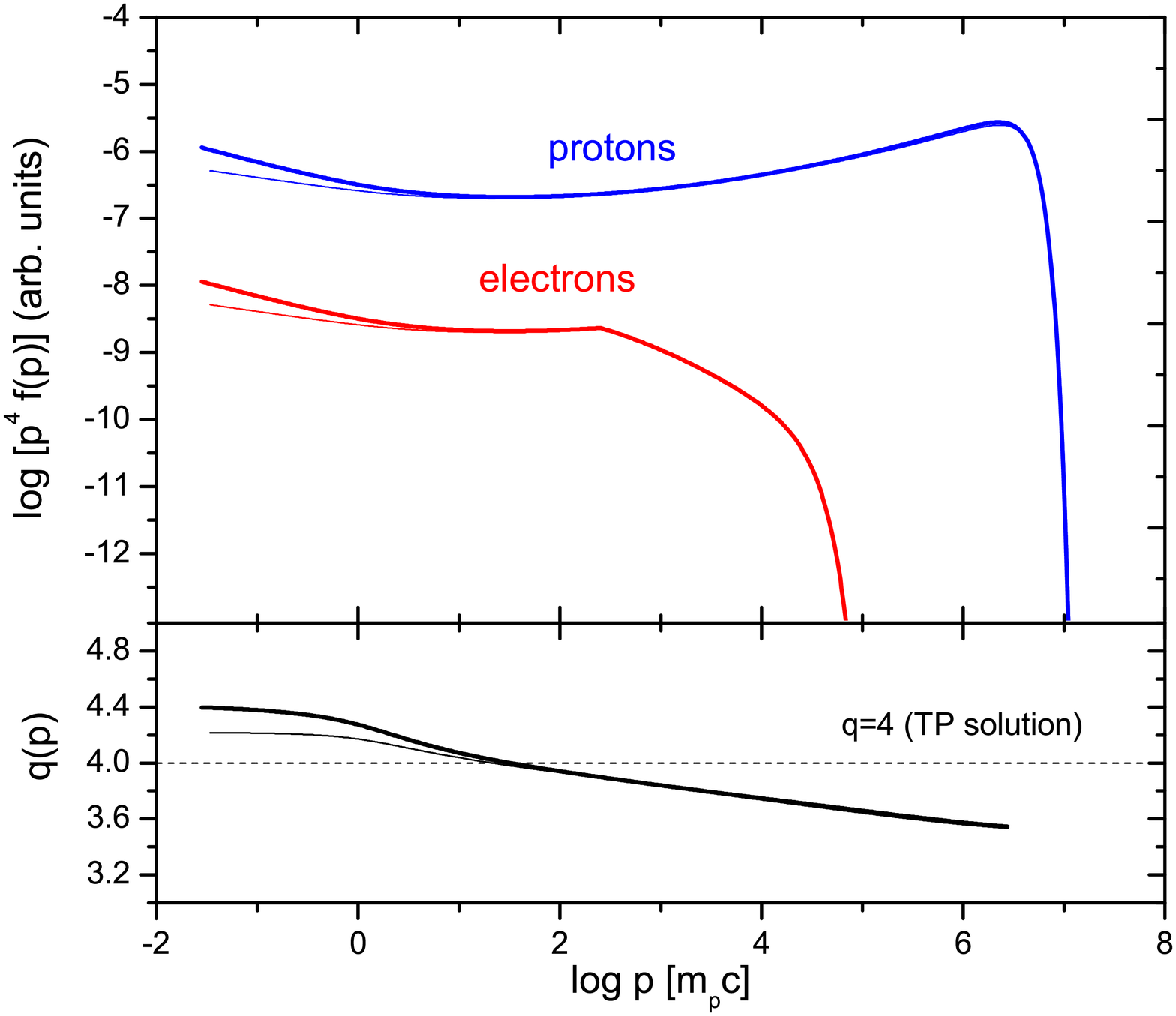}
    \caption{CR electron and proton non-thermal part of the spectra (top panel) at the age of 115 yr after explosion and proton
    spectrum slope $q(p)=-\frac{\mathrm{d} \ln{f_{\rm p,0}(p)}}{\mathrm{d} p}$ assuming particle distribution $\propto p^{-q}$ (bottom panel). Two representative models are shown, the most efficient NLDSA in Model A (thick
     solid line) and the least efficient NLDSA, amongst four models from
     Table~\ref{tab:models}, in Model D (thin solid line).}
    \label{fig:spectra}
\end{figure}

Our hydrodynamical approach does not 
self-consistently deal with the magnetic turbulence
and belonging higher order anisotropies which, according to \citet{bell11}, steepen the spectral index at quasi-perpendicular shocks. Due to this, we
fit SNR G1.9+0.3 radio spectra simply by using the
described non-linear acceleration model of Blasi and
assuming efficient acceleration (namely, acceleration efficiency up to $\eta \approx 10^{-3}$, to which corresponds $\xi$ between 3.3 and 3.45 in our simulations). Efficient acceleration is expected for such a young SNR and consistent with previous works \citep[e.g.,][]{ksenofontov10}. The electrons that mainly produce radiation by the synchrotron mechanism (in the amplified magnetic field $\sim$ 100 $\mu$G) at frequencies $\sim$ 1 GHz have energy $\sim$ 1 GeV and momentum $\sim m_pc$. At energies around 1 GeV, our energy spectra $N(E) \propto E^{-\gamma}$ of accelerated particles at CR modified shocks become softer, with the effective power-law index $\gamma$ around 2.2, giving the required spectral steepening of synchrotron spectra.

\begin{table*}
	\centering
	\caption{Model parameters for simulated radio spectra. In all models,
	ambient density was 0.02 cm$^{-3}$, $\alpha$ represents radio emission spectral index defined in Section 3, $\eta$ stands for the fraction of particles entering acceleration (as defined in Section 2.2) and $S_{1.425}$ and $S_{4.860}$ are respectively modelled flux densities at 1425 and 4860 MHz in Jy.}
	\label{tab:models}
	\begin{tabular}{ccccccccc} 
		\hline
		    & $\xi$ & $\zeta$ & $\eta$ [$\times 10^{-3}$] & $\alpha$ &  $S_{1.425}$  & $S_{4.860}$ & $R_{\rm tot}$ & $R_{\rm sub}$ \\
		\hline
	 Model A    & 3.30   & 0.40 & 1.1 & 0.619   &  0.933   & 0.440  & 12.0 &  3.2  \\     
	 Model B    & 3.35 & 0.33 & 0.9  & 0.605   &  0.931   & 0.446  & 11.7 &    3.3 \\     
	 Model C    & 3.40   & 0.24 &  0.7  & 0.591   &  0.928   & 0.452  & 11.3  &   3.4 \\     
	Model D    & 3.45  & 0.10 & 0.5 & 0.574   &  0.924   & 0.459  & 10.8  &    3.5  \\     
		\hline
	\end{tabular}
\end{table*}

From the fit of the synchrotron emission, we obtain value $2 \times 10^{-3}$ for the electron-to-proton ratio $K_{\rm ep}$, which is slightly lower than the value observed in the local CR spectra. Our value is in good agreement with values derived by other authors \citep[e.g.,][]{ksenofontov10, morlino12, slane14}, although \citet{yuan12} proposed a model in which $K_{\rm ep}=10^{-2}$ is a universal value that can fit all Galactic SNRs. As the dynamical role of electrons is negligible in the used model for NLDSA, this parameter acts as a scaling factor for the total radio emission and should not have any qualitative effects on radio evolution.

Two free parameters are of the utmost importance for the total radio flux in our simulations: injection parameter $\xi$, determining a fraction of particles $\eta$ entering the acceleration process and therefore global efficiency of NLDSA and MFA, and Caprioli's parameter $\zeta$, controlling heating of the plasma by non-linear damping of Alfv\'en waves and therefore being able to reduce the shock modification to some extent. 

As already said, efficient acceleration is necessary for the required spectral steepening of the radio spectra, but also $\xi$ values around 3.4 lead to very strong MFA and in turn can cause overestimate of the SNR total radio flux. Therefore, some damping is likely and a search through parameter space leads to the conclusion that, in our modelling, Caprioli's parameter $\zeta$ between 0.1 and 0.4 provides good fits. These values are also in agreement with \citet{kang13}, who arbitrarily set $ \zeta = 0.5$ in their four heuristic models of MFA in the precursor and which is, according to them, a reasonable estimate. We choose three best-fitting models, for injection parameter $\xi$ values 3.30, 3.35, 3.40 and 3.45, respectively, denoting them with Models A, B, C and D (Table~\ref{tab:models} and Fig.~\ref{fig:fit}). We allow four different scenarios, although producing similar radio spectra for particular  2008. epoch, in order to allow possible differences in simulations of radio flux temporal evolution, being the main goal of our paper. Fig. \ref{fig:fit} indicates that MOST measurement 0.97 $\pm$ 0.11 Jy should be taken with caution as probably being subject to significant measurement errors. In order to reproduce observed radio flux spectra from 2008, our simulations predict that the amplified magnetic field in downstream was then around $B_2=280\ \mu \rm{G}$. This value is in agreement with the value $\approx 230\ \mu \rm{G}$ inferred for G1.9+0.3 from equipartition calculations \citep{arbo12} and rapid variability in X-rays for notably older SNR RX J1713.72$-$3946, indicating amplification of the magnetic field by a factor of even more than 100 \citep{uchi07}. From the simulated 2500 yr of evolution, the obtained amplified magnetic field in the downstream is $B_2 \propto u_0^{0.76} \approx u_0^{3/4}$ (see Section~\ref{sec:mfa} for discussion).

\begin{figure}
	\includegraphics[width=\columnwidth]{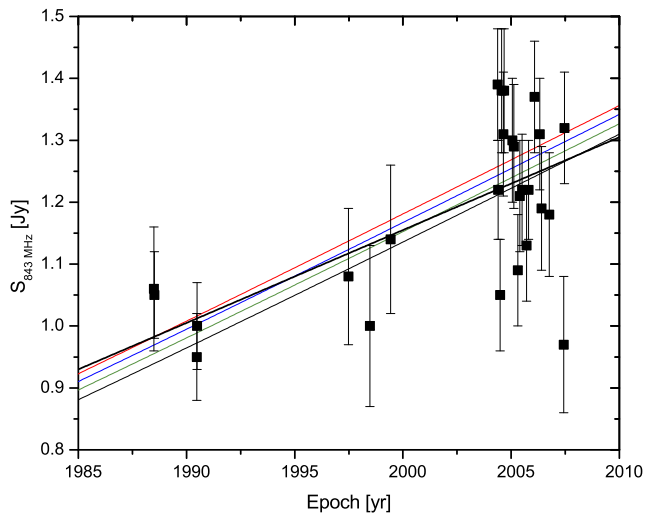}
    \caption{Observed and simulated evolution of the integrated radio flux density at 843 MHz. Black squares with error bars represent radio light curve for G1.9+0.3 from 1988 to 2007, observed by MOST  \citep{murphy08}. The black thick line shows a least-squares fit with gradient 0.015 Jy yr$^{-1}$ and a flux density of 1.23 Jy on 2005 January 1, originally obtained by \citet{murphy08}. Our simulations are shown with thin colour lines, namely Model A (red), Model B (blue), Model C (green) and Model D (black),
    showing similar average flux gradient $\approx$ 0.017 Jy yr$^{-1}$.}
\label{fig:radio-e}
\end{figure}

Fig.~\ref{fig:spectra} clearly shows that particle spectrum of accelerated particles changes from that predicted by standard linear
DSA, which reads $\propto p^{-4}$ in momentum and corresponds to energy distribution $\propto E^{-2}$ in a relativistic regime. Similarly to the 
cut-off for the electrons (Section~\ref{sec:radio}), we parametrize
the turnover of protons also by multiplying their distribution by an 
exponential factor, in a form suggested by \citet{lee2012}.
For sufficiently large momenta, the electron distribution function additionally deviates from the spectrum predicted by NLDSA, as a result of synchrotron losses.

As we mentioned before, a radio light curve for G1.9+0.3 based on 25 epochs
of observation with the MOST is available \citep{murphy08}.
These observations were taken with the same instrument, at constant frequency (843 MHz) and comparable resolutions (43 $\times$ 91 or 43 $\times$ 95 arcsec$^2$). Therefore, we run our numerical simulations
for different model parameters shown in Table \ref{tab:models} and synthesize the total radio flux density at frequency 843 MHz during the 
period from 1985 until 2010, in order to make comparison with 
 observations (Fig. \ref{fig:radio-e}). We obtained the average 
 flux gradient of around 0.017 $\rm{Jy} \,
 \rm{yr}^{-1}$ during this period (1.8 per cent yr$^{-1}$), which is in a very good agreement with a least-squares fit of MOST observations which gives 0.015 $\rm{Jy} \,\rm{yr}^{-1}$ (1.22$\pm^{0.24}_{0.16}$ per cent yr$^{-1}$) and also very close to the estimate of $\sim$2 per cent yr$^{-1}$ made by \citet{green08}, based on observations from a range of instruments, compiled from the literature. During the simulated part of the free expansion phase, derived dependence of the radio flux density is $S_{\nu} \propto D^{2.675}$, corresponding to radio surface brightness dependence $\Sigma_{\nu} \propto D^{0.675}$, while \citet{bif04} derive steeper dependence $\Sigma_{\nu} \propto D$ in free expansion. Interestingly, X-ray flux brightening of G1.9+0.3, measured by \citet{carlton11}, is also close to simulated and observed radio values, namely 1.7 $\pm$ 1.0 per cent yr$^{-1}$.

Three 11 cm flux density measurements from the Effelsberg
100-m radio telescope clearly show a strong flux density increase of
G1.9+0.3 for more than 30 yr. These measurements are respectively:
0.44$\pm$0.05 Jy from epoch 1983.48 \citep{reich84, furst90},
0.61$\pm$0.02 Jy from 2008.56 and 0.65$\pm$0.02 Jy from 
2016.72 (private communication, courtesy of Dr. Wolfgang Reich). 
The resulting rate of flux change for Effelsberg 
data is therefore around 0.006 Jy yr$^{-1}$, while
our model predicts rate 0.008 Jy yr$^{-1}$ at frequency 2695 MHz, 
for the corresponding period. Frequency independent flux expansion rate for Effelsberg data $\approx$1.4 per cent yr$^{-1}$ agrees well with
other measurements and our simulations.

We are also able to deduce the FS expansion rate from our simulations, for the evolution period from 1985 until 2010, roughly covering available observations. Our value of 0.9 per cent yr$^{-1}$ is in good agreement with expansion rate measurements $\approx$0.65 per cent yr$^{-1}$, from VLA observations \citep{green08}, and 0.64 $\pm$ 0.05 per cent yr$^{-1}$ \citep{carlton11}, obtained by comparing \emph{Chandra} X-ray images.

\begin{figure}
	\includegraphics[width=\columnwidth]{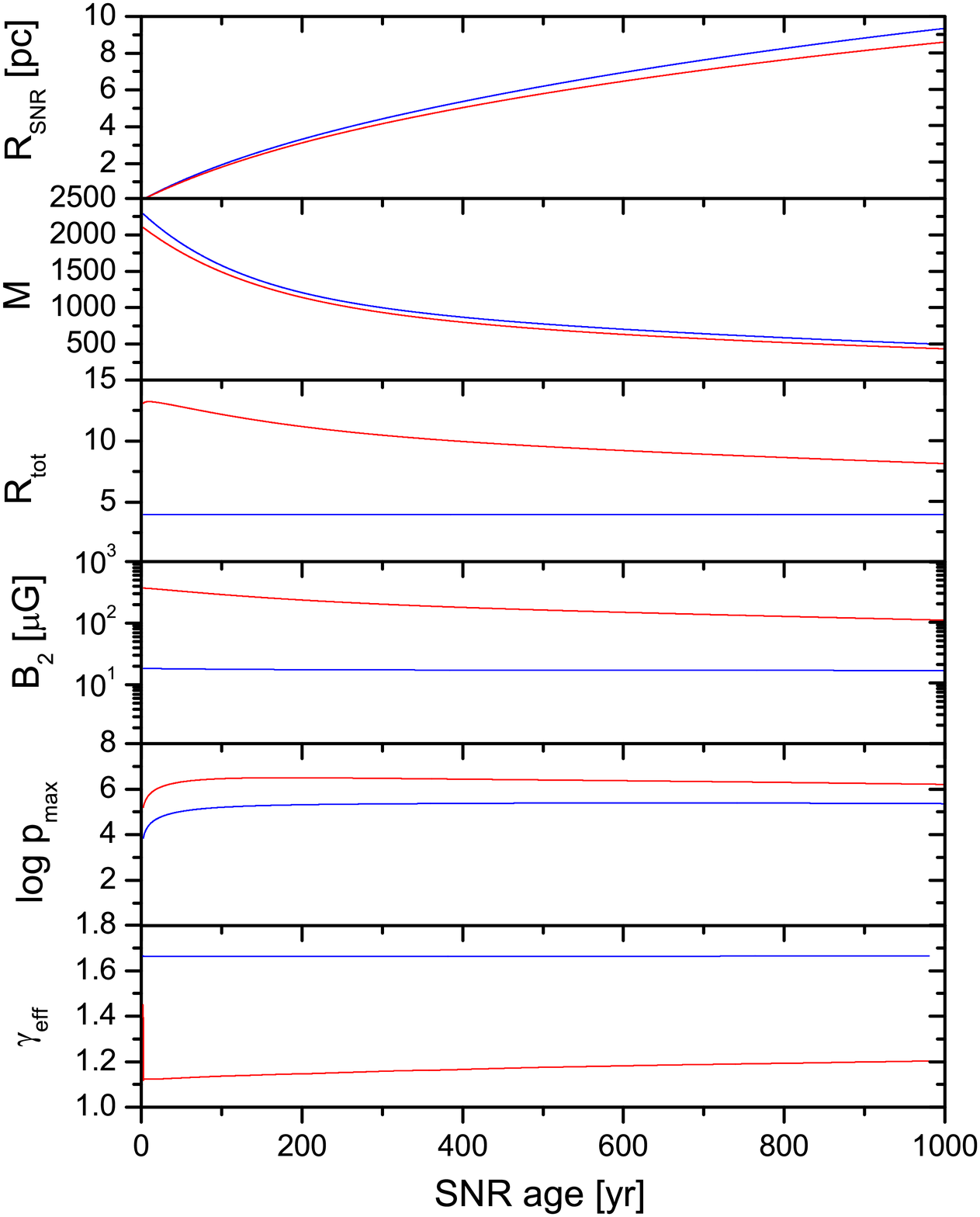}
    \caption{Evolution parameters of G1.9+0.3 for a period of 1000 yr. The blue line represents TP DSA, namely for $\xi = 4.8$, and the red line
    represents NLDSA with Model A parameters (the most efficient 
    acceleration among models listed in Table \ref{tab:models}).}
   \label{fig:evol-compare}
\end{figure}

In Fig. \ref{fig:evol-compare}, we plot time evolution of different characteristics of SNR for period of 1000 yr and compare the evolution corresponding to efficient DSA, including MFA, with a TP case.
The red line represents DSA with pronounced non-linear effects and strongly modified shock (parameters from Model A), while the blue line represents SNR evolution with injection parameter $\xi = 4.8$ leading to almost negligible amount of accelerated CRs. The total compression in the TP case reduces to 4 and magnetic field downstream is amplified only due to gas compression, which is far below the value required for the observed radio emission.

\begin{figure}
\includegraphics[width=\columnwidth]{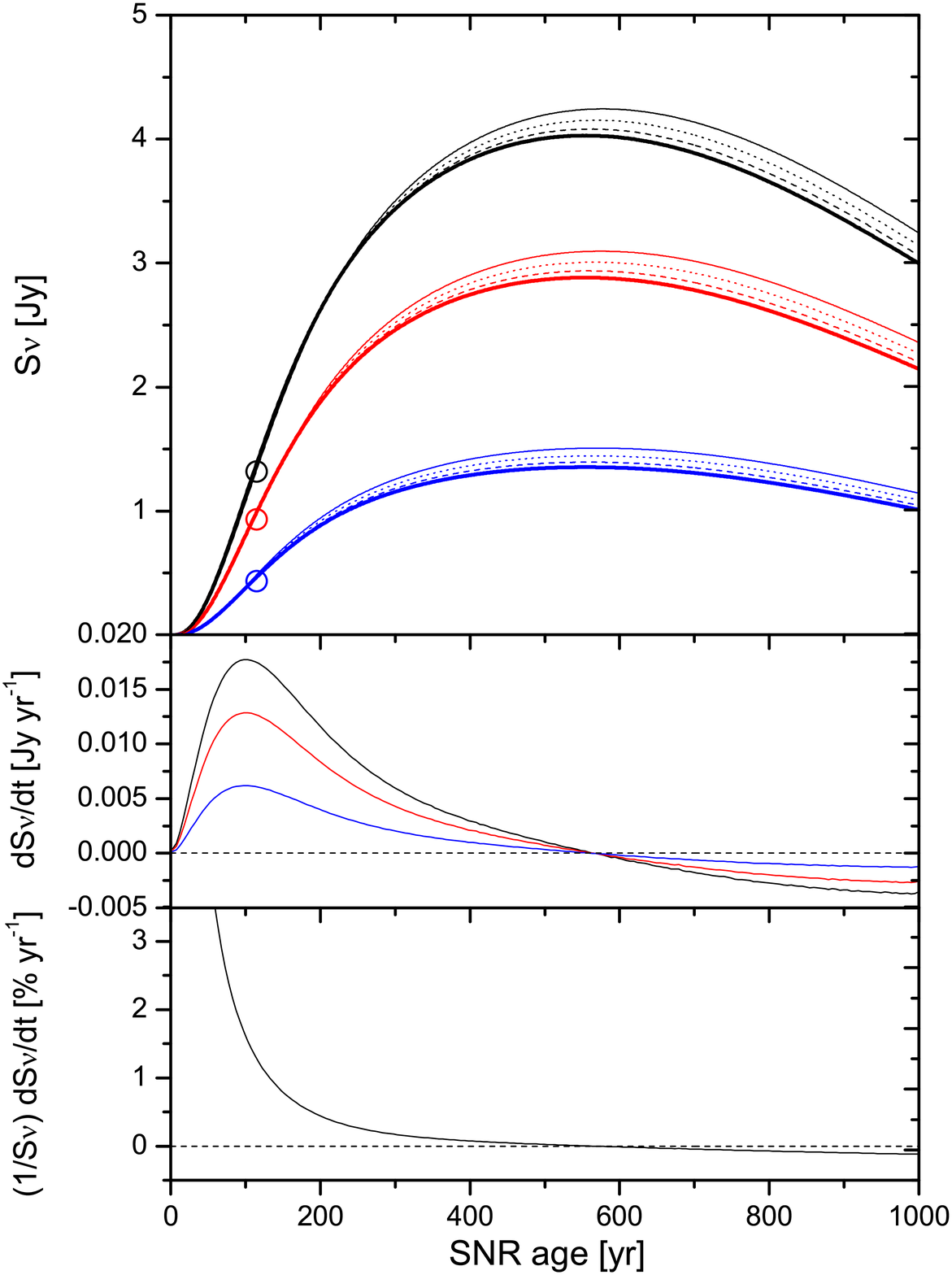}
\caption{Simulated radio evolution of G1.9+0.3 during 1000 yr after SN explosion. The upper panel shows radio flux density evolution at three frequencies: 843 MHz (black lines), 1425 MHz (red) and 4860 MHz (blue).
Four models are shown for each frequency: Model A (thick solid line), Model B (dashed), Model C (dotted) and Model D (thin solid). Open circles denote radio flux densities observed in 2008. The middle panel shows the corresponding rate of change of flux density $\frac{\mathrm{d} S_{\nu}}{\mathrm{d} t}$ in Jy yr$^{-1}$, averaged for each frequency and line colours having the same meaning as in the upper panel. The lower panel shows the annual flux density increase, $\frac{1}{S_{\nu}}\frac{\mathrm{d} S_{\nu}}{\mathrm{d} t}$, in per cent yr$^{-1}$, which is independent of frequency. The maximum flux density is predicted around 600 yr, which is before the shock sweeps ISM mass equal to $3 M_{\sun}$, roughly marking the end of the free expansion phase in our simulation ($\approx 1700$ yr).}
   \label{fig:evol-S1}
\end{figure}

\begin{figure}
\includegraphics[width=\columnwidth]{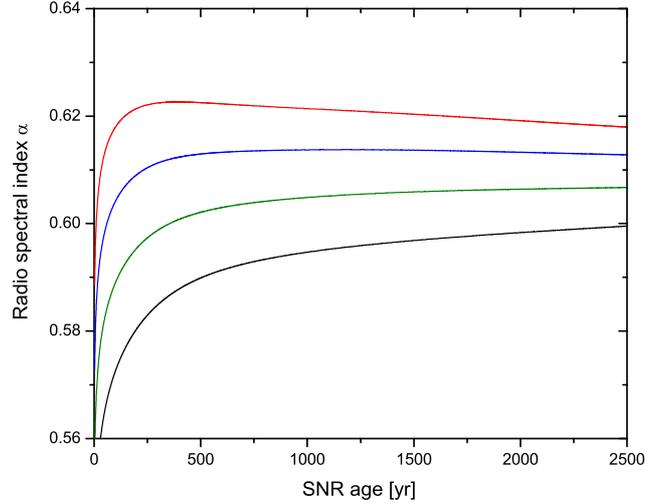}
\caption{Simulated evolution of the radio spectral index $\alpha$, defined so that the flux density depends on frequency as $S_{\nu} \propto \nu^{-\alpha}$. Spectral index evolution is model dependent and due to this we denote different scenarios with colours, namely Model A (red line), Model B (blue), Model C (green) and Model D (black).}
   \label{fig:alpha}
\end{figure}

\begin{figure}
	\includegraphics[width=\columnwidth]{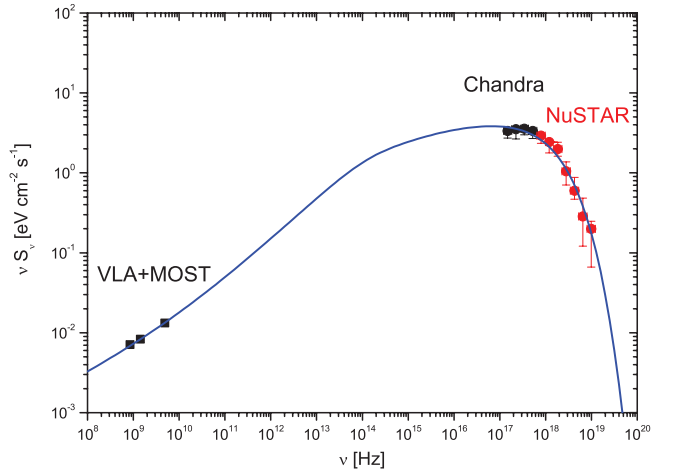}
    \caption{Spatially integrated synchrotron flux density as a function of frequency from the radio to the X-ray domain. The observational data are, respectively: radio from VLA \citep{green08} and MOST \citep{murphy08}; X-ray fluxes are taken from \citet{yang16}, originally observed with \emph{Chandra} \citep{reyn08} and NuSTAR \citep{zog15} telescopes.}
\label{fig:xray}
\end{figure}

Being the main purpose of this paper, we simulate the time dependence of
flux densities $S_{\nu}$ of the radio synchrotron emission at different frequencies with its corresponding rate of change $\dot{S_{\nu}}$ in Jy yr$^{-1}$ and frequency-independent fractional change $\frac{\dot{S_{\nu}}}{S_{\nu}}$ in percentage yr$^{-1}$ (Fig.~\ref{fig:evol-S1}). Our model predicts increasing radio emission from G1.9+0.3 during the part of the free expansion phase, reaching its maximum value around the age of 600 yr and then decreasing during the later free expansion and Sedov phase. For the determined ambient density of 0.02 cm$^{-3}$, radius around 11.3 pc marks the end of the free expansion (ejecta dominated) phase, which we assume to be when the swept-up mass is $M_{\rm{sw}}=3M_{\rm{ej}}$.
In our simulations, this happens $\approx$ 1700 yr after the SN explosion. Model also predicts maximum radio flux densities $\sim$ 4.3, 3.1 and 1.5 Jy, respectively for frequencies 843, 1425 and 4860 MHz, being around three times higher than the present values. It can be inferred from Fig. \ref{fig:evol-S1} that models with higher injection parameter $\xi$ (less efficient acceleration) give slightly higher flux densities close to maximum, for a chosen frequency. This is mainly due to the increasing efficiency of MFA in Models from A to D (see Table \ref{tab:models}), linked with the parameter $\zeta$. As for the rate of flux density change during time (in Jy yr$^{-1}$), our model suggests a maximum at around $t_{\rm{SNR}} \sim 100$ yr followed by gradual slowing down of flux increase, until it starts to decline. Interestingly, available measurements of radio light curve for G1.9+0.3 roughly coincide with this maximum \citep{green08, murphy08, carlton11}, meaning they probably contain the fastest ever flux change for this SNR.

Simulations also give insight into the radio spectral index $\alpha$ evolution (Fig. \ref{fig:alpha}), reflecting the evolution of the spectrum of accelerated electrons with energies around $\sim$ GeV. Evolution starts from the values close to $\alpha=0.5$, corresponding to the TP DSA solution, reaches maximum value (the steepest radio spectra) and then slowly decreases, making spectra shallower. Evolutionary tracks for radio spectral index were obtained by implementing models from Table \ref{tab:models} and they seem strongly model dependent. Higher injection efficiency (lower $\xi$ parameter) naturally leads to higher value for $\alpha$ in maximum but also this maximum is reached earlier in the SNR lifetime. This is in good agreement with a considerable amount of observational evidence for steep spectral indices of SNRs being $\sim$ 1000 yr old or even few times older, as radio spectral index slowly decreases after reaching maximum steepness. The greatest number of evolved SNRs have spectral indices in the interval 0.5 $\leq$ $\alpha$ $\leq$ 0.6, as the DSA predicts \citep{urosevic14}.

\begin{figure}
	\includegraphics[width=\columnwidth]{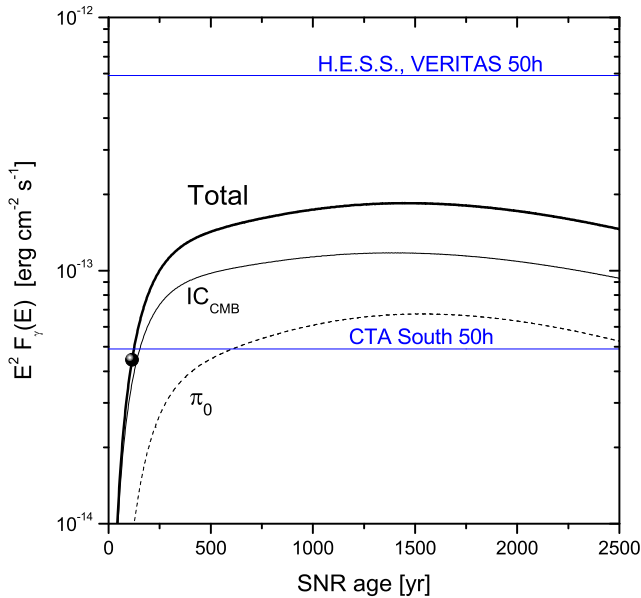}
    \caption{Evolution of spatially integrated 4 TeV gamma-ray emission produced by pion decay (dashed line) and IC computed for the CMB photon field (thin solid line). The thick solid line shows the sum of these two contributions. The blue lines represent the corresponding flux sensitivities of H.E.S.S., VERITAS and CTA gamma telescopes. The black ball represents the current evolutionary status.}
\label{fig:gamma}
\end{figure}

Synchrotron emission spans from the radio to the X-ray band.
\citet{reyn08} concluded that X-ray emission from G1.9+0.3 appears to be
purely synchrotron radiation. We neglect X-ray emission due to thermal bremsstrahlung and synthesize the synchrotron spectrum up to the highest energies by using an electron spectrum obtained in our simulations (Fig. \ref{fig:xray}) and adopted to fit observations, as described in Section~\ref{sec:radio}. Magnetic field $B_2=280\ \mu$G inferred from our simulations for the SNR age of $t_0=115$ yr in 2008 gives the position of the break in the electron spectrum around $p_b \approx 10^{3} m_p c$ (Equation \ref{eq:Eb}). The best-fitting maximum value for the steepening $\delta$ is 0.5, implemented through our \lq toy\rq~model (Section \ref{sec:radio}) as uniformly growing function starting from momenta slightly before $p_b$, to ensure physically more consistent and smoother transition from the uncooled to the cooled regime instead of a sharp break in the electron spectrum\footnote{\citet{tanaka08} mention possible detection of this synchrotron break in optical/infrared wavelengths as additional argument in favour of strong MFA.}. The synchrotron spectrum in Fig. \ref{fig:xray} is obtained only for Model A and it is obtained for the maximum electron cut-off momentum $p_{\rm e,max} = 10^{4.5} m_{\rm p} c$ and the corresponding energy $E_{\rm e,max}= 27$ TeV, obtained by assuming Bohm diffusion. Spectra for the remaining three models (Table \ref{tab:models}) were omitted because they are very similar. Also, modelled spectra of G1.9+0.3 reveals concave-up curvature at millimetre and sub-millimetre  wavelengths. This is expected and indeed observed by the \emph{Planck}\footnote{A project of the European Space Agency (ESA). It observes the sky in nine frequency bands covering 30--857 GHz with high sensitivity and angular resolution.} telescope in the radio continuum of another young SNR Cas A \citep{onic15}. After  investigating alternative explanations of the observed curvature, \citet{onic15} agree that non-linear effects of particle acceleration are mainly responsible for high-frequency curvature in the radio spectrum. 
 
Following the approach described in Section~\ref{sec:gamma}, we evolve the SNR during the 2500 yr and calculate its 4 TeV gamma-ray emission produced by pion decay and IC computed for the CMB photon field (Fig. \ref{fig:gamma}). We chose to model 4 TeV emission intentionally, as the highest CTA sensitivity is expected around this energy\footnote{Reference: CTA energy flux sensitivity, \url{www.cta-observatory.org}}. For the present SNR age, the expected total TeV gamma-ray emission is $4.4 \times 10^{-14}~\rm{erg}~\rm{cm}^{-2}~\rm{s}^{-1}$. Such a flux is too low for possible H.E.S.S.\footnote{H.E.S.S.: Preliminary sensitivity curves for H.E.S.S.-I (stereo reconstruction), adapted from \citet{holler15}} or VERITAS\footnote{Very Energetic Radiation Imaging Telescope Array System  (VERITAS): public specifications webpage \url{veritas.sao.arizona.edu/about-veritas-mainmenu-81/veritas-specifications-mainmenu-111}} detection in $\sim$ 50 h, more than one order of magnitude below their sensitivities. On the other hand, present value from our model is slightly below the predicted CTA (Southern Site) sensitivity limit of $4.9 \times 10^{-14}~\rm{erg}~\rm{cm}^{-2}~\rm{s}^{-1}$ around TeV energies, but expected to reach this limit within a decade or so. The pion decay flux is only about 1/4 of the IC gamma-ray flux, probably as a result of the low ambient gas density. The maximum TeV gamma-ray flux is predicted to occur around the end of the free expansion phase, at the age of 1500 yr, and it reaches $1.8 \times 10^{-13}~\rm{erg}~\rm{cm}^{-2}~\rm{s}^{-1}$. This value is still below the sensitivity limit of H.E.S.S., but probably visible in TeV gamma-rays by future instruments, including the CTA.
 
 However, more advanced broad-band modelling of G1.9+0.3 is out of the scope of this paper. The X-ray part of the spectrum and gamma-ray emission evolution are given above only in an illustrative way, to check how our model fits with observations in domains other than radio. More rigorous numerical treatment of synchrotron losses will be necessary in order to obtain evolution of the emission at energies higher than radio. We reserve a detailed modelling of SNR evolutionary tracks at different wavelengths for future work.

\section{Conclusion}

The peculiar nature of radio evolution of the youngest known Galactic 
SNR G1.9+0.3 is modelled by using Blasi non-linear kinetic theory of CR acceleration in SNRs coupled with 3D hydrodynamics, simultaneously solved with the {\scriptsize PLUTO} code. We assume this SNR originated from a Type Ia supernova (SN) explosion located near the GC, with explosion energy 10$^{51}$ erg and ejecta mass 1.4 M$_{\sun}$.
Hydrodynamic equations in the {\scriptsize PLUTO} code were adopted to use the space and time-dependent adiabatic index in order to account for the presence of energetic particles, making the fluid more compressible. Our modelling and analysis leads to the following essential results.

(i) From our 3D hydrodynamic simulations of SNR evolution, including a 
deceleration of FS by the ambient medium and due to back reaction of CRs, we estimate the current age of G1.9+0.3 SNR to be
slightly over 120 yr, expanding in an ambient density of 0.02 cm$^{−3}$.

(ii) Efficient acceleration is necessary in order to explain
observed spectral steepening of the radio spectra. Namely,
observations are well fitted for injection parameter $\xi$ between 3.45 and 3.30, corresponding to an acceleration efficiency $\eta =($0.5--1.1$) \times 10^{-3}$ and magnetic field amplified more than 50 times from the assumed ambient value.

(iii) Following our models, it can be concluded that radio emission increasing brightness is a common property of young SNRs. Our model gives the average 843 MHz flux increase gradient during a 20-yr period of around 0.017 $\rm{Jy} \,\rm{yr}^{-1}$ (1.8 per cent yr$^{-1}$), which is in a very good agreement with MOST observations and also with other available observations from a range of instruments, compiled from the literature.
Simulations give the average 2695 MHz flux gradient of 0.008 Jy yr$^{-1}$ during the past 30 yr, being in a good agreement with Effelsberg measurements.

(iv) Numerical model predicts increasing radio emission from G1.9+0.3 during the free expansion phase, reaching its maximum value around 
the age of 600 yr and then decreasing during late free expansion and beginning of Sedov phase around 1700 yr after the SN explosion. Interestingly, it seems that we are currently witnessing approximately the fastest radio emission increase than it will ever be.

(v) The radio brightness will grow according to prediction given in this paper, until its maximum flux densities of $\sim$ 4.3, 3.1 and 1.5 Jy, respectively, for frequencies 843, 1425 and 4860 MHz, being around three times higher than the present day values.

(vi) The steep radio spectral index (steeper than linear DSA prediction of $\alpha=0.5$) for young SNRs is explained only by means of efficient NLDSA and accompanying strong MFA. The radio spectral index also shows qualitatively similar evolution as the radio flux, it reaches the steepest value $\alpha_{\rm{max}}$ and then becomes shallower (trending towards the value of 0.5). Higher injection efficiency $\eta$ leads to higher $\alpha_{\rm{max}}$ but also causes this value to be reached earlier in the SNR history. However, the temporal evolution of the radio spectral index turns out to be very sensitive to model parameters $\xi$ and $\zeta$.

(vii) We implement a simple \lq toy\rq~model for the synthesis of a broader synchrotron
spectrum from radio to X-ray, by using the electron spectrum obtained in our simulations. This spectrum is modified in post-processing by introducing a break in the electron spectrum, to account for synchrotron losses and modelled X-ray emission fit well the \emph{Chandra} and NuSTAR measurements.
It agrees well with models of spectra containing more consistent, numerical calculation of synchrotron losses.

(viii) We also implement approximative model of gamma-ray emission coming from the SNR. We inspect time evolution of the total gamma-ray flux and conclude that it may be visible in TeV gamma-rays by future instruments, including the CTA. Model predicts increasing TeV gamma-ray emission during the entire free expansion phase, reaching the maximum value of $1.8 \times 10^{-13}~\rm{erg}~\rm{cm}^{-2}~\rm{s}^{-1}$ at the age of around 1500 yr.

Our model enabled us to make important conclusions about the present and predictions about the future properties of radio emission from the youngest known Galactic SNR. We want to emphasize that, although the presented model contains robust implementation, all provided quantitative estimates should be taken with caution. Besides our limited knowledge in
physical descriptions of particle acceleration and SNR evolution, a
significant number of model parameters still remain weakly constrained.

Models of radio evolutionary tracks can be of the utmost importance for 
the future observers working on powerful radio telescopes
such as ALMA\footnote{The Atacama Large Millimeter/submillimeter Array} 
and SKA\footnote{The Square Kilometre Array}. These types of modelling can provide important information about the evolutionary stage of 
SNRs, as well as to characterize the physical conditions in the shocks where the relativistic particles are accelerated.

\section*{Acknowledgements}

This work is part of project no. 176005 \lq Emission nebulae:
structure and evolution\rq\,supported by the Ministry of Education, Science, and Technological Development of the Republic of Serbia. Numerical simulations were run on the PARADOX-IV supercomputing facility at the Scientific Computing Laboratory of the Institute of Physics Belgrade, supported in part by the Ministry of Education, Science and Technological Development of the Republic of Serbia under projects no. ON171017 and OI1611005. 
Simulations were also run on cluster Jason, belonging to Automated Reasoning Group (ARGO) based at the Department of Computer Science, Faculty of Mathematics, University of Belgrade.
I thank the anonymous referee for the 
very constructive suggestions on this manuscript.
I would like to thank D. Uro\v sevi\' c and B. Arbutina for introducing me
into this exciting field, continually supporting and encouraging me, but also for careful reviewing and editing of typescript.
I acknowledge the hospitality of the Osservatorio Astronomico di
Palermo where part of this work was carried out, special thanks to Salvatore Orlando and Marco Miceli for their illuminating contributions to this project and also for reading the manuscript. I'm grateful to Gilles Ferrand for extremely helpful discussions, advices and help during this work and coding. Brian Reville provided valuable discussion on different approaches in SNR modelling. I'm indebted to Tara Murphy for kindly providing of currently available MOST radio flux densities, to Dr. Wolfgang Reich for providing Effelsberg radio data and to Rui-Zhi Yang and Leonid Ksenofontov for sharing the X-ray data for G1.9+0.3 and useful explanations. {\scriptsize PLUTO} is developed at the Turin Astronomical Observatory in collaboration with the Department of Physics of the Turin University.






\label{lastpage}
\end{document}